\documentclass[final,3p,times, twocolumn]{elsarticle}

\usepackage[T1]{fontenc}
\usepackage[latin9]{inputenc}
\usepackage{array}
\usepackage{multirow}
\usepackage{amsmath}
\usepackage{amssymb}
\usepackage{graphicx}
\usepackage{grffile}
\usepackage{wasysym}
\usepackage{verbatim}
\usepackage{multicol}
\usepackage{float}

\makeatletter

\providecommand{\tabularnewline}{\\}

\usepackage{chemmacros}
\newcommand{\stst}{\standardstate}

\makeatother

\begin{document}

\title{Rigorous Analysis of Non-Ideal Solubility of Sodium and Copper Chlorides in Water Vapor Using Pitzer-Pabalan Model}

\author{K. A. Velizhanin}
\ead{kirill@lanl.gov}
\address{Theoretical Division, Los Alamos National Laboratory, Los Alamos, NM 87545, USA}

\author{C. D. Alcorn, A. A. Migdisov}
\address{Earth and Environmental Sciences, Los Alamos National Laboratory, Los Alamos, NM 87545, USA}

\author{R. P. Currier}
\address{Chemistry Division, Los Alamos National Laboratory, Los Alamos, NM 87545, USA}

\date{\today}
\begin{abstract}
Gaseous mixtures of water vapor and
neutral molecules of salt (e.g., NaCl, CuCl etc.) can be highly non-ideal
due to the strong attractive interaction between salt and water molecules. In particular, this can result in high solubility
of salts in water vapor and a strong dependence of solubility on vapor
pressure. The analysis of salt solubility in water vapor can be
done using the Pitzer-Pabalan model, which is based on the thermodynamic
theory of imperfect gases. The original Pitzer-Pabalan work demonstrated that one can
reproduce experimental data for ${\rm NaCl}$ solubility in vapor. No analysis was performed on the reliability
of their original fits, which we believe has contributed to the lack of applications of the Pitzer-Pabalan model despite the apparent
success of the original paper. In this work, we report recent
progress in developing a rigorous fitting procedure to parameterize
the Pitzer-Pabalan model using experimental data. Specifically, we
performed fitting of the experimental results obtained elsewhere for
NaCl and CuCl. We investigate the degree of underfitting/overfitting
and the sensitivity of the fitting quality to variations in the resulting
fitting parameters. The results, as represented by the
thermodynamic parameters describing the energetics of formation of
salt-bearing water clusters, were successfully benchmarked against
Gaussian 16 {\it ab initio} quantum chemistry calculations. The resulting rigorous fitting procedure presented here can now be applied to other systems. 
\end{abstract}
\maketitle

\section{Introduction}

It is an experimental fact that a gaseous mixture of water vapor and
neutral salt molecules (e.g., ${\rm NaCl}$, ${\rm CuCl}$ etc)
can be highly non-ideal with respect to the salt due to strong attractive interaction
between the salt and water molecules in water vapor \citep{Galobardes-1981-363,Pitzer-1983-1120,Bischoff-1986-1437,Pitzer-1986-1445,Armellini-1993-123,Archibald-2002-1611,Pokrovski-2005-657,Pokrovski-2008-345,Migdisov-2013-123}.
This means that the chemical potential of the salt in such equilibrium mixtures is significantly lower than it would have been, at the
same concentration of salt, if water and salt molecules were not interacting.
In particular, the concentration of salt in water vapor, brought to
equilibrium with bulk solid salt, can be many orders of magnitude
higher than the concentration of a saturated anhydrous gas in equilibrium
with the solid salt. Accurate quantitative understanding of this non-ideality
and, in particular, of this last phenomenon is very important for
understanding various geological processes and for various industrial applications
ranging from desalination to corrosion.

Experimental data are often represented as the concentration
of salt in vapor, in equilibrium with bulk crystalline salt, measured over
a range of temperatures and pressures. To make this data useful
for general applications, one can generally use two approaches. The first approach is to fit the experimental data, for example,
the salt concentration as a function of vapor pressure at a given
temperature, by a certain empirical expression whose functional form
is motivated more by ease of use than by underlying physics \citep{Archibald-2002-1611,Migdisov-2013-123,Migdisov-2014-33}.
This approach is expected to work fine when interpolating experimental
results. However, using this method to extrapolate experimental data
to temperature and pressures beyond the ranges accessible in experiment,
or to describe solubility in more complex systems (e.g., containing
several gases where the density of gas mixture is not fully controlled
by water vapor), can be questionable. The second, semi-empirical, approach is to fit experimental data with
functional forms based on the underlying physics. For ${\rm NaCl}$
solubility in water vapor, this was done by Pitzer and Pabalan
\citep{Pitzer-1986-1445}. In short, their model was based on treating
tightly-bound clusters made of water and salt molecules as separate
molecular species. Introduction of the Gibbs free energy of formation,
and its temperature derivatives, for such clusters naturally yields
the concentration of salt in vapor as a function of temperature and
pressure of water vapor in equilibrium. Since these Gibbs free energies
are properties of a specific cluster, and do not depend on concentration
of these clusters, this approach in principle allows for extrapolation
to higher or lower water vapor pressures. Furthermore, since the functional form of the semi-empirical
expressions for the salt concentration are based on correct physical
principles, they automatically produce correct asymptotic behavior,
e.g., in the case of very large or very low water vapor pressure. Extrapolation
to higher and lower temperatures, beyond those covered in the experimental data, is also a possibility. The downside of the semi-empirical approach is that it yields more complicated expressions than those in the
empirical approach.

In their seminal paper \citep{Pitzer-1986-1445}, Pitzer and Pabalan
successfully fit experimental results for vapor solubility of ${\rm NaCl}$.
However, some details of the fitting procedure they used remained
unexplained. For example, a single fitting variable was assigned to
the enthalpy of hydration of a salt-bearing water cluster for the
first three hydration steps beyond the anhydrous ${\rm NaCl}$ molecule.
Another independent variable was assigned to the enthalpy of the next
three hydration steps and so on. It is unknown if there was any physical
motivation behind this assignment choice. Furthermore, it is unclear if (and unlikely that)
this parameterization is universal enough to be able to accommodate
other salts (e.g., ${\rm CuCl}$). Finally, it is not clear whether the dependence
of the enthalpy of hydration on the number of hydration steps Pitzer
and Pabalan used was not overfitting the data, i.e., if the temperature
and pressure ranges of experimental data, as well as the data quality
(e.g, data scattering) allowed them to uniquely constrain the values
of all the independent model parameters. It is a possibility
that the values of enthalpies of formation they listed was only one
choice of a variety of possible choices that can fit the experimental
data equally well. Reliability of the extrapolation beyond the experimental
data would be questionable if this last scenario is realized. On the
other hand, underfitting, where the fitting function was not flexible
enough to properly fit experimental data, is also possible.

We believe that these uncertainties relating to the
original work by Pitzer and Pabalan have contributed to the absence of publications where experimentally
observed salt solubility in vapor were analyzed with their model. In this work, we perform a data fitting of ${\rm NaCl}$ and
${\rm CuCl}$ solubility in water vapor, obtained in previous publications
\citep{Bischoff-1986-1437,Pitzer-1986-1445,Archibald-2002-1611,Migdisov-2014-33},
analyzing rigorously (i) the degree of underfitting and overfitting,
(ii) sensitivity of the fit quality to variations of fitting parameters,
and (iii) what happens if some of the previously fixed
model parameters are now treated as fitting variables. The last scenario
is relevant in a situation where, for example, the thermodynamics
of sublimation of salt is not available from thermochemical tables.
Furthermore, to see whether extracted thermodynamic parameters are
reasonable, we also compare them to those obtained from quantum
chemical calculations performed with the Gaussian 16 software package,
and also to data available from thermochemical tables.

The paper is organized as follows. The Pitzer-Pabalan model for salt
solubility in water vapor is introduced in Sec. \ref{sec:theory}.
To set the stage for the analysis of solubility, we first discuss
the thermodynamics of pure-water and pure-salt clusters in Secs. \ref{sec:H2O_therm}
and \ref{sec:X_therm}, respectively. Solubilities of ${\rm NaCl}$
and ${\rm CuCl}$ in vapor are then rigorously analyzed in Secs. \ref{sec:NaCl_solubility}
and \ref{sec:CuCl_solubility}, respectively. Extracted thermodynamic parameters are discussed in Sec. \ref{sec:Discussion}. Our conclusions are found in Sec. \ref{sec:Conclusion}.

\section{Pitzer-Pabalan Model\label{sec:theory}}

In the imperfect gas theory \citep{Hill-1986-StatTherm, McQuarrie-2000-StatMech}, the interaction between molecules in a sufficiently dilute multicomponent molecular gas leads to the virial equation of state, where the pressure is represented as a multivariable power series in fugacities of the molecular species. It can be shown, that in the case of a strong attractive interaction between the molecules, which results in the formation of tightly bound molecular clusters, the virial coefficients of such an expansion are the equilibrium constants of formation of such clusters \citep{Hill-1986-StatTherm}. The virial expansion can then be reinterpreted as one corresponding to an ideal mixture of ideal gases of molecular clusters with the total pressure given by $P=\sum_i P_i$, where $P_i =c_i RT$ is the partial pressure of an ideal gas of clusters of the same type (e.g., the same stoichiometry). The cluster molarities $c_i$ are obtained from the equilibrium constants of cluster formation and are thus determined by the strength of interaction of molecules within clusters. In other words, strong attractive interactions between molecules in a dilute gas can be treated indirectly by introducing chemical-like reactions in an otherwise ideal gas. For example, the attractive interaction between two water molecules in the water vapor can be accounted for indirectly by considering the reaction
\begin{equation}
{\rm H_{2}O}+{\rm H_{2}O}\rightarrow{\rm (H_{2}O)_{2}},\label{eq:H2Odim_form}
\end{equation}
where a new species is introduced - a tightly bound water dimer ${\rm (H_{2}O)_{2}}$. This approach belongs to the class of the so called quasi-chemical approximations \citep{Hill-1986-StatTherm}, another example being the Hayden-O'Connell theory of dimerization \citep{Hayden-1975-209}. The validity of the approach is based on a bound cluster being a well-defined entity, i.e., the system has to be sufficiently dilute to keep clusters spatially separated from each other. For example, the virial expansion for water vapor fails when approaching the vapor-liquid phase boundary where the density is too high to consider clusters spatially isolated and non-interacting, see Eq. (\ref{eq:P_H2O_summed}) and the corresponding discussion below.

Pitzer and Pabalan used the imperfect gas theory together with the assumption of tightly bound clusters to describe salt solubility in water vapor \citep{Pitzer-1986-1445}. They considered a gas of water and salt (e.g., ${\rm NaCl}$ or ${\rm CuCl}$, denoted as ${\rm X}$ below) molecules in equilibrium with crystalline salt in the absence of liquid water. In the spirit of the quasi-chemical approximation, the interactions between water and salt molecules were described by the following set of reactions:
\begin{subequations}
\begin{equation}
{\rm X(cr)}\rightarrow{\rm X},\label{eq:react_sublim}
\end{equation}
\begin{equation}
{\rm H_{2}O}+{{\rm X}_{ m}\text{:}({\rm H_{2}O})_{n-1}}\rightarrow{{\rm X}_{m}}\text{:}({\rm H_{2}O})_{ n},\label{eq:react_h2o}
\end{equation}
\begin{equation}
{\rm X}+{{\rm X}_{m-1}\text{:}({\rm H_{2}O})_{n}}\rightarrow{{\rm X}_{m}}{\rm \text{:}(H_{2}O)}_{ n}.\label{eq:react_salt}
\end{equation}
\end{subequations}
Here, Eq. (\ref{eq:react_sublim}) describes the extraction of the
single molecule of ${\rm X}$ from the solid crystalline salt ${\rm X(cr)}$
into an ideal gas of single salt molecules. Typically, the gaseous
salt molecule would be denoted by ${\rm X(g)}$ in literature, but we omit ${\rm (g)}$ for brevity. Thus, for example, ${\rm H_{2}O}$
and ${\rm X}$ represent ideal gas of single water and salt molecules,
respectively. A cluster made of $m$ salt molecules and $n$ water
molecules is denoted by ${{\rm X}_{m}}\text{:}({{\rm H_{2}O}})_{{ n}}$.
Accordingly, an addition of a single gaseous water or salt molecule
to such a cluster is represented by Eqs. (\ref{eq:react_h2o}) and
(\ref{eq:react_salt}), respectively.

\begin{figure*}[!htb]
\centering
\includegraphics[width=0.65\paperwidth]{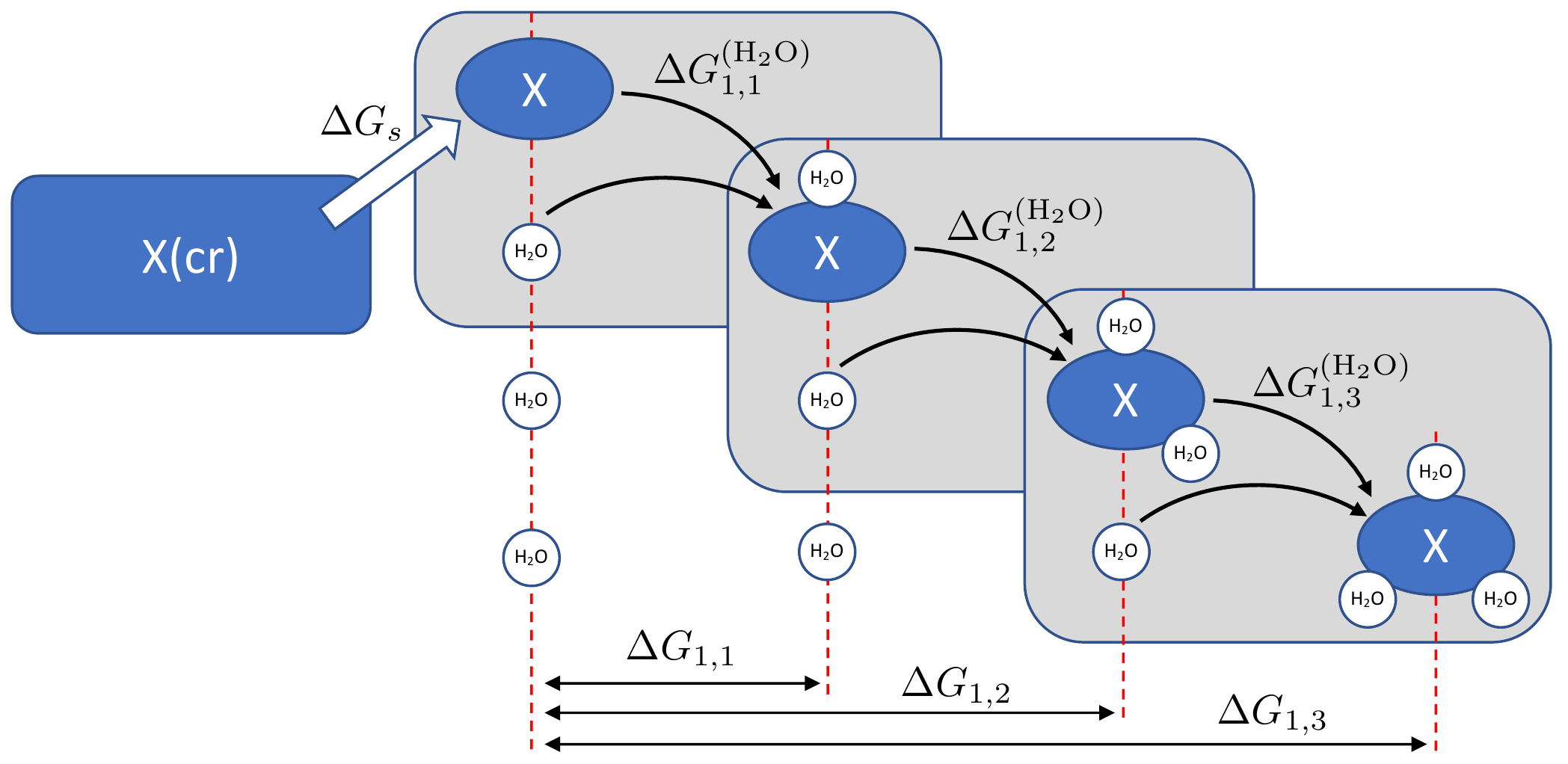}\caption{\label{fig:Schematics}Schematics of successive formation of salt-bearing
water cluster.}

\end{figure*}

\begin{figure*}[!htb]
\centering
\includegraphics[width=0.65\paperwidth]{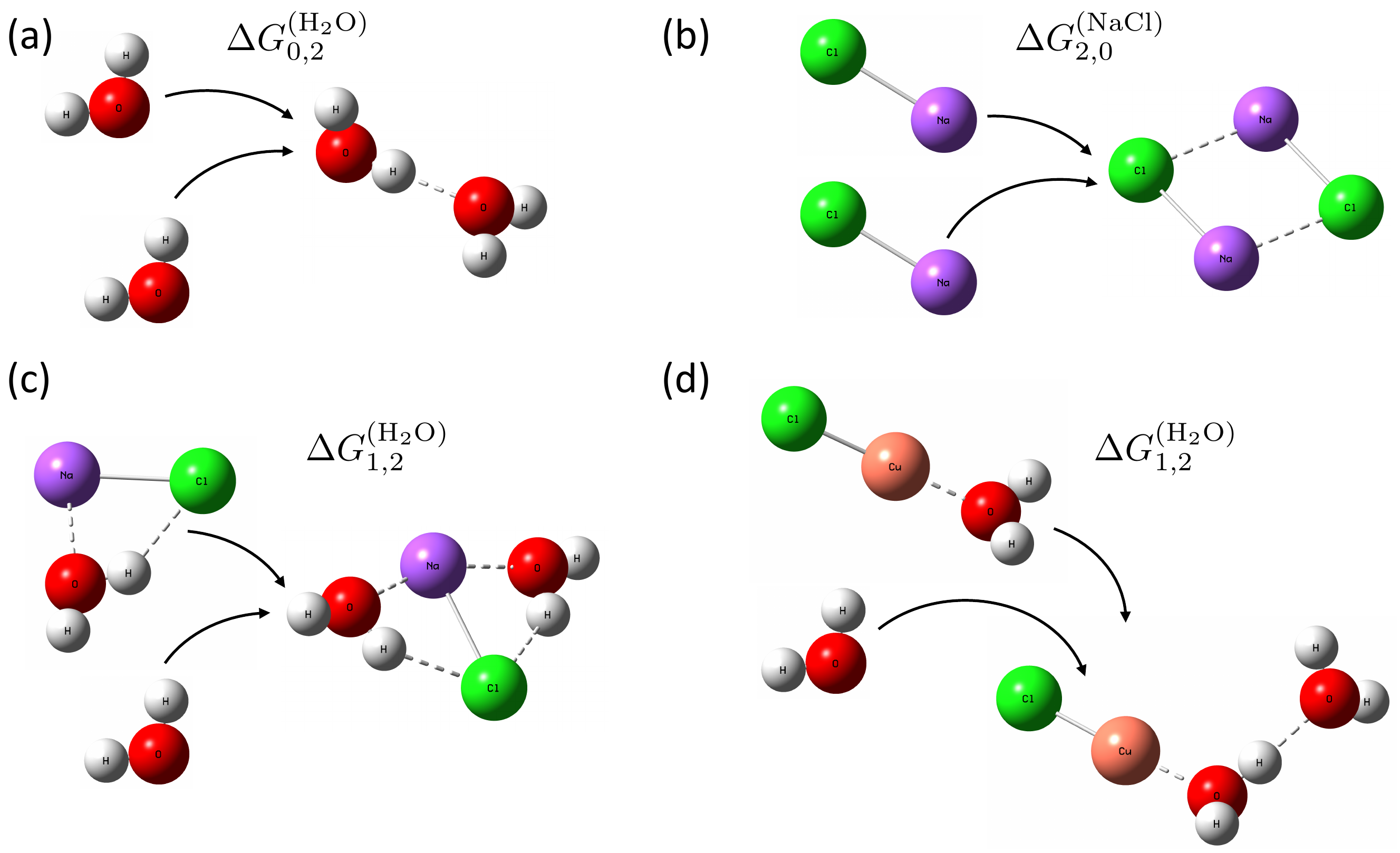}\caption{\label{fig:atomistics}Several representative cluster formation processes.
For each process, the change of the Gibbs free energy in the process
is written. Cluster geometries are obtained from the quantum chemical
calculations with the Gaussian 16 software package \citep{Gaussian16}. }

\end{figure*}

The changes in partial molar Gibbs free energies corresponding to Eqs. (\ref{eq:react_sublim})-(\ref{eq:react_salt}) read, respectively, as
\begin{subequations}
\begin{equation}
\Delta G_{s}=G_{1,0}-G_{cr},\label{eq:dG_s}
\end{equation}
\begin{equation}
\Delta G_{m,n}^{({\rm H_{2}O})}=G_{m,n}-G_{0,1}-G_{m,n-1},\label{eq:dG_H2O}
\end{equation}
\begin{equation}
\Delta G_{m,n}^{(X)}=G_{m,n}-G_{1,0}-G_{m-1,n},\label{eq:dG_X}
\end{equation}
\end{subequations}
Here, the partial molar Gibbs free energy (or molar chemical potential) of an ideal gas of
clusters ${{\rm X}_{m}\text{:}({\rm H_{2}O})_{n}}$ (or in shorthand $(m,n)$-clusters)
is denoted by $G_{m,n}$. The partial molar Gibbs free energy of the solid
crystalline salt ${\rm X(cr)}$ is denoted by $G_{cr}$. The Gibbs free
energy of sublimation is denoted by $\Delta G_{s}$. The change of
the molar Gibbs free energy in the process where a single water or
salt molecule is added to a cluster to produce the $(m,n)$-cluster
is denoted by $\Delta G_{m,n}^{({\rm H_{2}O})}$ or $\Delta G_{m,n}^{({\rm X})}$,
respectively. In what follows, $\Delta G_{m,n}^{({\rm H_{2}O})}$
will be referred to as the Gibbs free energy of hydration. The set
of reactions leading to the successive formation of a cluster containing
a single molecule of salt, and associated changes in Gibbs free energies,
are depicted in Figure \ref{fig:Schematics}.
Some examples of formation of clusters are shown in Figure \ref{fig:atomistics}.
The partial molar Gibbs free energy of formation of a $(m,n)$-cluster out of $m$
and $n$ gaseous molecules of salt and water, respectively, is denoted
by
\begin{equation}
\Delta G_{m,n}=G_{m,n}-mG_{1,0}-nG_{0,1},\label{eq:dG_mn}
\end{equation}
where the notation is again chosen so that the pair of the subscripts
in the l.h.s. denote the product species - the $(m,n)$-cluster. Using
Eqs. (\ref{eq:dG_H2O}) and (\ref{eq:dG_X}), $\Delta G_{m,n}$ can
be written as a sum of energy changes corresponding to successive
additions of single water and salt molecules to a cluster. For example,
\begin{equation}
\Delta G_{1,n}=\sum_{i=1}^{n}\Delta G_{1,i}^{({\rm H_{2}O})},\label{eq:dG_1n}
\end{equation}
as is illustrated in Figure \ref{fig:Schematics}. This allows us to write
$\Delta G_{1,n}^{({\rm H_{2}O})}=\Delta G_{1,n}-\Delta G_{1,n-1}$,
that is $\Delta G_{1,n}^{({\rm H_{2}O})}$ can be considered a derivative
of $\Delta G_{1,n}$ with respect to $n$, evaluated by the finite
difference method, i.e., $\Delta G_{1,n}^{({\rm H_{2}O})}\sim\frac{d(\Delta G_{1,n})}{dn}$.

Since the gas of clusters ${{\rm X}_{m}\text{:}({\rm H_{2}O})_{n}}$ is assumed
ideal, its pressure is given by
\begin{equation}
P_{m,n}/P^{\circ}=e^{(G_{m,n}-G_{m,n}^{\circ})/RT},\label{eq:P_mn}
\end{equation}
where $R$ is the gas constant, $P^{\circ}=1\,{\rm bar}$ - standard
pressure, and $G_{m,n}^{\circ}(T)=G_{m,n}(T,P^{\circ})$ is the standard
(i.e., at standard pressure, temperature is arbitrary) partial molar Gibbs
free energy of the ideal gas of $(m,n)$-clusters. In what follows,
we will generally omit writing $T,P$ arguments unless required. In
equilibrium, the l.h.s of Eqs. (\ref{eq:dG_s})-(\ref{eq:dG_X}) vanish
and and one can demonstrate that $G_{m,n}=mG_{cr}+nG_{0,1}$ expectedly
holds. Substituting this into the Gibbs free energy change in Eq.
(\ref{eq:P_mn}) we obtain
\begin{align}
G_{m,n}-G_{m,n}^{\circ}&=m(G_{cr}-G_{cr}^{\circ})+n(G_{0,1}-G_{0,1}^{\circ})\nonumber \\
&-m\Delta G_{s}^{\circ}-\Delta G_{m,n}^{\circ},\label{eq:G_mn_Gmn_st}
\end{align}
where $\Delta G_{m,n}^{\circ}$ is given by Eq. (\ref{eq:dG_mn}),
taken at the standard pressure. The fugacity of water vapor, or more simply water fugacity, is introduced as
\begin{equation}
f=P^{\circ}e^{(G_{0,1}-G_{0,1}^{\circ})/RT}.\label{eq:f_H2O}
\end{equation}
The relative activity of salt is $a_{s}=e^{(G_{cr}-G_{cr}^{\circ})/RT}$. Eq. (\ref{eq:G_mn_Gmn_st}) then becomes
\begin{align}
G_{m,n}-G_{m,n}^{\circ}&=mRT\ln a_{s}+nRT\ln\frac{f}{P^{\circ}}\nonumber\\
&-m\Delta G_{s}^{\circ}-\Delta G_{m,n}^{\circ}.
\end{align}
Since the crystalline salt is only weakly compressible, one can assume the salt molar volume independent of pressure and apply the Poynting method to estimate the variation of $G_{cr}(T,P)$ with
pressure \citep{Dahm-2015-Chemical}. More specifically, $G_{cr}(T,P)-G_{cr}^{\circ}(T)\approx V_{s}^{\circ}(P-P^{\circ})$,
where $P$ is the total pressure in the system and $V_{s}^{\circ}$
is the molar volume of the crystalline salt at the standard pressure.
Typical experimental pressures are up to $P\sim200\,{\rm bar}$ at
characteristic temperatures $T\sim600\,{\rm K}$ \citep{Migdisov-2014-33}.
As an example, for ${\rm NaCl}$ with an ambient density of 2.16 ${\rm g/cm^{3}}$
and molar mass of 58.4 $g/mol$, the relative activity is $a_{s}\approx1.1$. The ambient bulk modulus of ${\rm NaCl}$ is $\sim 24$ GPa \citep{Hofmester-1997-5835}, which leads to relative changes of molar volume with pressure on the level of $200\:{\rm bar} / 24\:{\rm GPa}\sim10^{-3}$, thus justifying the assumption of pressure-independent molar volume.
The results are very similar for ${\rm CuCl}$. In what follows, we
set $a_{s}=1$. Therefore, Eq. (\ref{eq:P_mn}) can be rewritten as
\begin{equation}
P_{m,n}/P^{\circ}=e^{-\Delta\tilde{G}_{m,n}^{\circ}/RT}\left(\frac{f}{P^{\circ}}\right)^{n},\label{eq:P_nm}
\end{equation}
where $\Delta\tilde{G}_{m,n}^{\circ}=m\Delta G_{s}^{\circ}+\Delta G_{m,n}^{\circ}$
- the change of the standard molar Gibbs free energy in a process
where $m$ salt molecules are extracted from the crystalline bulk
and combined with $n$ gaseous molecules of water to form the $(m,n)$-cluster.
The resulting total pressure is
\begin{equation}
P/P^{\circ}=\sideset{}{^{'}}\sum_{m,n}e^{-\Delta\tilde{G}_{m,n}^{\circ}/RT}\left(\frac{f}{P^{\circ}}\right)^{n},
\end{equation}
where the primed sum means that summation is done over all pairs $m,n\geq0$
that satisfy $m+n>0$. In particular, we can write
\begin{align}
P/P^{\circ}&=\sum_{n=1}^{\infty}e^{-\Delta G_{0,n}^{\circ}/RT}\left(\frac{f}{P^{\circ}}\right)^{n}\nonumber\\
&+\sum_{n=0}^{\infty}e^{-\Delta\tilde{G}_{1,n}^{\circ}/RT}\left(\frac{f}{P^{\circ}}\right)^{n}+...,\label{eq:P_tot}
\end{align}
where the first r.h.s term corresponds to imperfect water vapor with
no salt in it. Instead of explicitly performing the summation in this term, we use a Python
implementation of IAPWS-95 - an accurate equation of state (EOS) for
water vapor \citep{Wagner-2002-387,iapws_python}. The second r.h.s.
term represents $(1,n)$-clusters. We will demonstrate below, in Sec. \ref{subsec:Sublimation},
that only these two terms are important for conditions considered
in this work, or in other words, contributions from clusters bearing
two or more salt molecules are negligible. Under these conditions,
the partial pressure of the salt-bearing clusters is
\begin{equation}
P_{{\rm X}}/P^{\circ}=\sum_{n=0}^{\infty}e^{-\Delta\tilde{G}_{1,n}^{\circ}/RT}\left(\frac{f}{P^{\circ}}\right)^{n}.\label{eq:Px}
\end{equation}
Here, each summation term is proportional to the concentration of
the cluster of the corresponding size. The mean number of water molecules
in a cluster can, therefore, be calculated as
\begin{equation}
\langle n\rangle=\frac{\sum_{n=0}^{\infty}ne^{-\Delta\tilde{G}_{1,n}^{\circ}/RT}\left(\frac{f}{P^{\circ}}\right)^{n}}{\sum_{n=0}^{\infty}e^{-\Delta\tilde{G}_{1,n}^{\circ}/RT}\left(\frac{f}{P^{\circ}}\right)^{n}}=\frac{\partial\ln P_{{\rm X}}}{\partial\ln f}.\label{eq:mean_n}
\end{equation}

Values of $\Delta\tilde{G}_{1,n}^{\circ}$ are functions of temperature, and in order to obtain them from fits to experimental data,
we need to approximate them with a few-parameter functional form.
As in Ref. \citep{Pitzer-1986-1445}, we assume that the constant-pressure
heat capacities, defined by the second derivative of $\Delta\tilde{G}_{1,n}^{\circ}$
with respect to temperature, are temperature-independent constants.
This results in
\begin{align}
\Delta\tilde{G}_{1,n}^{\circ}(T)&=\Delta\tilde{H}_{1,n}^{\stst}-T\Delta\tilde{S}_{1,n}^{\stst}\nonumber\\
&-T\Delta\tilde{C}_{1,n}^{\stst}\left[\ln\frac{T}{T_{{\rm ref}}}+\frac{T_{{\rm ref}}}{T}-1\right],\label{eq:dG_1n_T}
\end{align}
where the three parameters $\Delta\tilde{H}_{1,n}^{\stst}$, $\Delta\tilde{S}_{1,n}^{\stst}$
and $\Delta\tilde{C}_{1,n}^{\stst}$ are the molar enthalpy, entropy
and constant-pressure heat capacity changes for the process of formation
of $(1,n)$-cluster out of $n$ gaseous water molecules and a salt
molecule, extracted from bulk, evaluated at the standard pressure $P^{\circ}$ and the reference temperature $T_{{\rm ref}}$. To facilitate
comparison with results of Ref. \citep{Pitzer-1986-1445}, the reference
temperature is taken $T_{{\rm ref}}=500\,{\rm K}$. We emphasize again
that superscript $\circ$ stands for the standard state at the standard
pressure ($P^{\circ}=1\,{\rm bar}$), but arbitrary temperature, whereas superscript $\stst$ also
sets the temperature to be $T=T_{{\rm ref}}$. In this paper, $C$ denotes the constant-pressure molar heat capacity. We,
therefore, omit $P$ in the usual notation $C_{P}$ for brevity. Expressions
analogous to Eq. (\ref{eq:dG_1n_T}) will also be used to represent
the temperature dependence of $\Delta G_{s}^{\circ}$, $\Delta G_{1,n}^{\circ}$,
$\Delta G_{1,n}^{\circ({\rm X})}$ and $\Delta G_{1,n}^{\circ({\rm H_{2}O})}$.
The accuracy of assuming a temperature-independent heat capacity in Eq. (\ref{eq:dG_1n_T}) will be discussed in Sec. \ref{subsec:dC_Tindep}.

\section{Thermodynamics of $\left({\rm H_{2}O}\right)_{\rm n}$ cluster formation\label{sec:H2O_therm}}

In this and the following section, we discuss the thermodynamics of forming pure-water and pure-salt clusters, respectively. Primarily,
we do this because we expect that the thermodynamics of forming salt-bearing water clusters will be somewhere ``in between''
those of pure-water and pure-salt. Secondary, since good quality experimental
data is available for thermodynamics of pure clusters, it can serve
as a benchmark for the reliability of the quantum chemistry-based
calculations. 

\begin{figure*}[!htb]
\centering
\includegraphics[width=0.8\paperwidth]{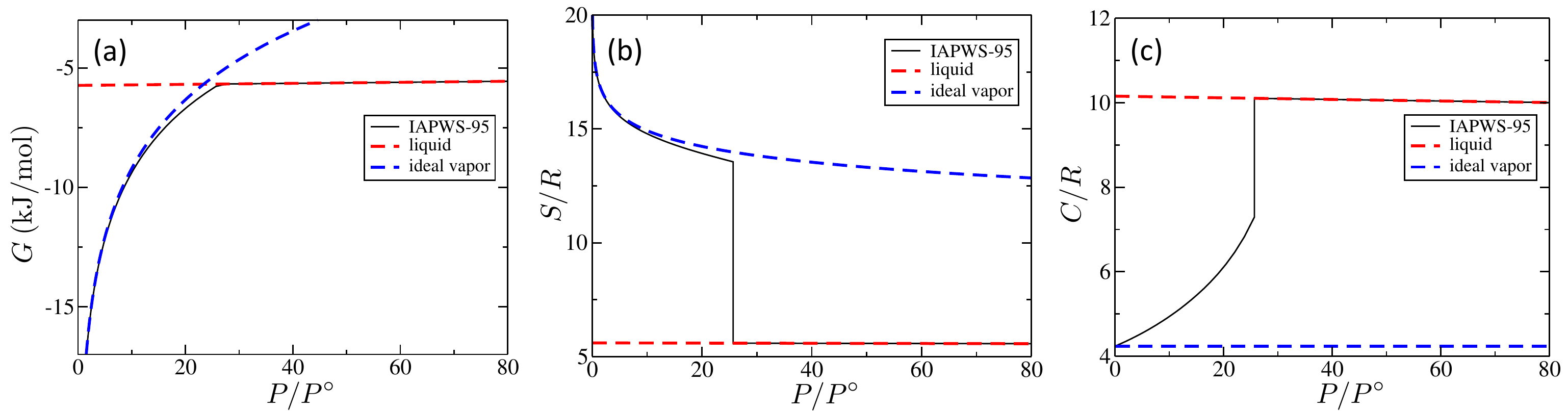}\caption{\label{fig:water_GSCp}
Molar Gibbs free energy, entropy and constant-pressure heat capacity for water (IAPWS-95 EOS) are plotted as functions of
pressure at $T=T_{{\rm ref}}$.}
\end{figure*}

First, consider the process of forming large water clusters,
Eq. (\ref{eq:react_h2o}) at $m=0$ and $n\gg1$. It turns out that
the complete thermodynamic description of such a process can be extracted
from a complete EOS that describes both vapor and liquid water.
The black line in Figure \ref{fig:water_GSCp} shows the dependence
of the molar Gibbs free energy for water at $T=T_{{\rm ref}}$, evaluated
from the IAPWS-95 EOS \citep{Wagner-2002-387,iapws_python}.

\begin{figure}[!htb]
\centering
\includegraphics[width=0.38\paperwidth]{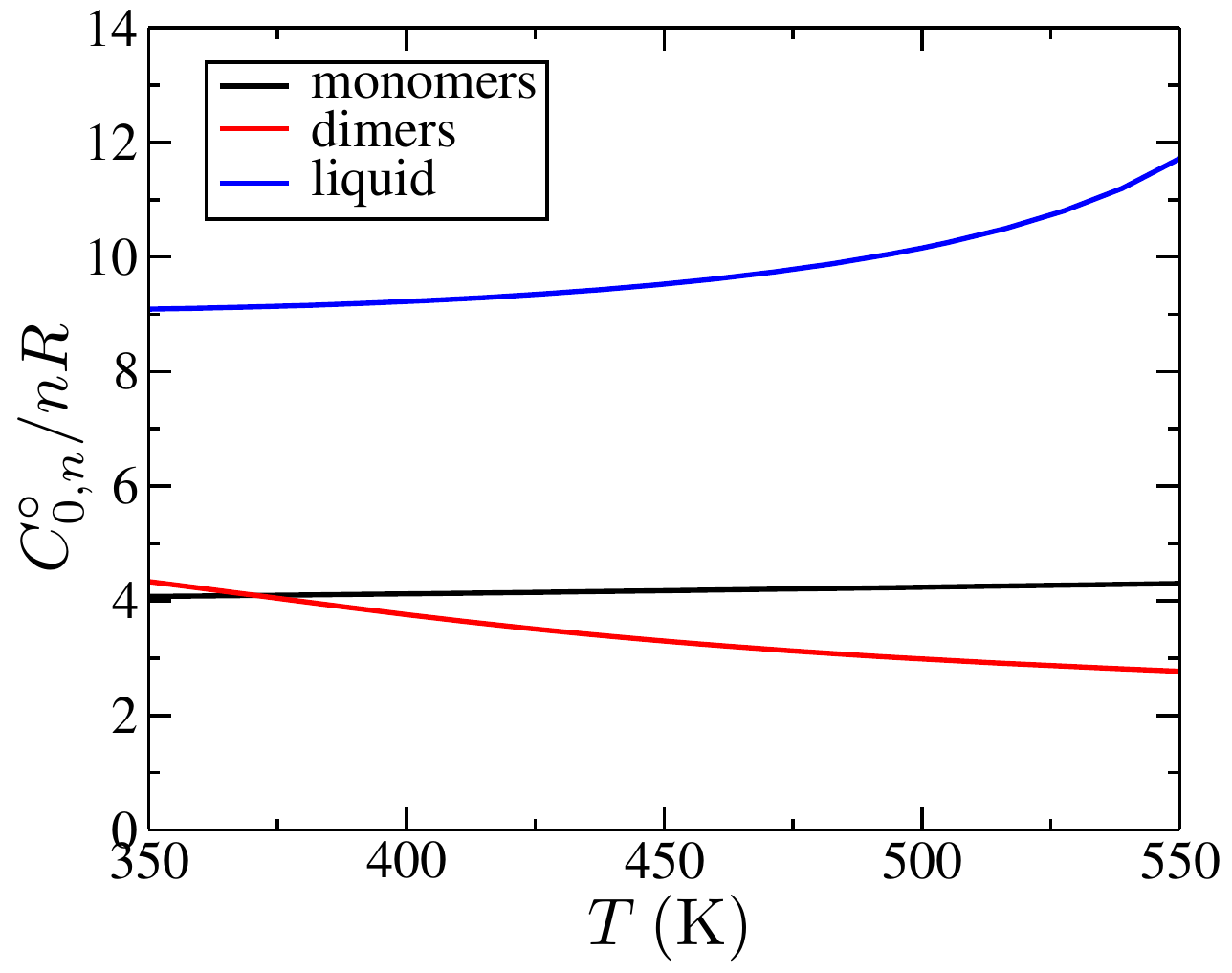}\caption{\label{fig:water_CvCp}
Molar constant-pressure heat capacities extracted
from IAPWS-95 for ideal gas of monomers ($n=1$), dimers ($n=2$)
and very large clusters ($n=\infty$) at standard pressure. Heat capacities
are normalized per mole of water molecules, $C_{0,n}^{\circ}/nR$.}
\end{figure}

At this temperature, liquid water is a thermodynamically stable phase
at $P\apprge28\,{\rm bar}$, with vapor being stable at lower pressures.
The vapor-liquid phase transition appears as a derivative discontinuity in the black line. To obtain the thermodynamic parameters for a process involving addition of a single water molecule to a large pure-water cluster,
we first consider the thermodynamics of the ideal vapor. According to
the first r.h.s term in Eq. (\ref{eq:P_tot}), the vapor becomes ideal
(i.e., no water clusters, just monomers) at infinitely low pressures.
In practice, we assume that the vapor is sufficiently ideal at some
low pressure $P'$ (e.g., $\sim10^{-3}\,{\rm bar}$). The partial molar Gibbs free
energy for the ideal vapor at arbitrary pressure $P$ is then evaluated
as
\begin{equation}
G_{0,1}(T,P)=G(T,P')+RT\ln P/P',\label{eq:G_P_Pp}
\end{equation}
where $G(T,P')$ is calculated directly from IAPWS-95 and other thermodynamic
state functions are obtained from $G_{0,1}(T,P)$ using thermodynamic
derivatives. The resulting Gibbs free energy, entropy and heat capacity
of the ideal vapor are plotted as dashed blue lines in Figure \ref{fig:water_GSCp}(a),
(b) and (c), respectively. That $P'$ was taken sufficiently low is
confirmed by, for example, panel (a) where the dashed blue line agrees
perfectly with IAPWS-95 at low pressures. Close to the boiling
point (e.g., $P\sim20\,{\rm bar}$), the Gibbs free energy of the
real vapor deviates from ideality due to the contribution of dimers
and larger clusters. As an illustration to using thermodynamic derivatives to obtain various thermodynamic parameters from Eq. (\ref{eq:G_P_Pp}), the dependence of constant-pressure heat capacity for the ideal vapor (i.e., the gas of monomers) at standard pressure, computed from the second derivative of $G_{0,1}$ with respect to temperature, is represented by the black line in Figure \ref{fig:water_CvCp}.

Now consider an ideal gas of very large salt-free water clusters
($m=0$, $n\gg1$). Each such cluster can be treated as a droplet
of liquid water \citep{Lemke-2008-3293,Machlin-2007-Aspects,Frenkel-1939-538} and therefore the partial Gibbs free energy for a mole of large $(0,n)$-clusters can be written as
\begin{equation}
G_{0,n}^{\circ}(T)\approx nG_{{\rm liq}}(T),\label{eq:nFliq}
\end{equation}
where $G_{{\rm liq}}(T)$ is the Gibbs free energy of a mole of
water molecules in the liquid phase.
We neglected the contributions of translation and rotation of the cluster as the whole, as well as the surface
energy term, in the r.h.s. of Eq. (\ref{eq:nFliq}), since
they are small compared to $nG_{{\rm liq}}$ for sufficiently large clusters \citep{Machlin-2007-Aspects}.
An important point is that since the imperfect gas theory assumes
that clusters do not interact directly, $G_{{\rm liq}}$ has to
be evaluated at zero pressure, and not at $P^{\circ}$.  Thermodynamic information for liquid water at zero pressure
is not directly available since the liquid water is not a thermodynamically stable phase at zero pressure. However, thermodynamic variables
are seen to vary slowly with pressure in Figure \ref{fig:water_GSCp}
in the liquid regime ($P\apprge28\,{\rm bar}$), which is the result of the very low compressibility of liquid water. We can thus extrapolate thermodynamic state functions, evaluated where the liquid phase is still stable, to zero pressure using, for example, a low-degree
polynomial regression. This approach yields thermodynamic properties of a metastable phase - the liquid water at zero pressure. The constant-pressure heat capacity of liquid water, extracted by this method and normalized to a mole of water molecules, is represented by the blue line in Figure \ref{fig:water_CvCp}.

\begin{figure*}[!htb]
\centering
\includegraphics[width=0.8\paperwidth]{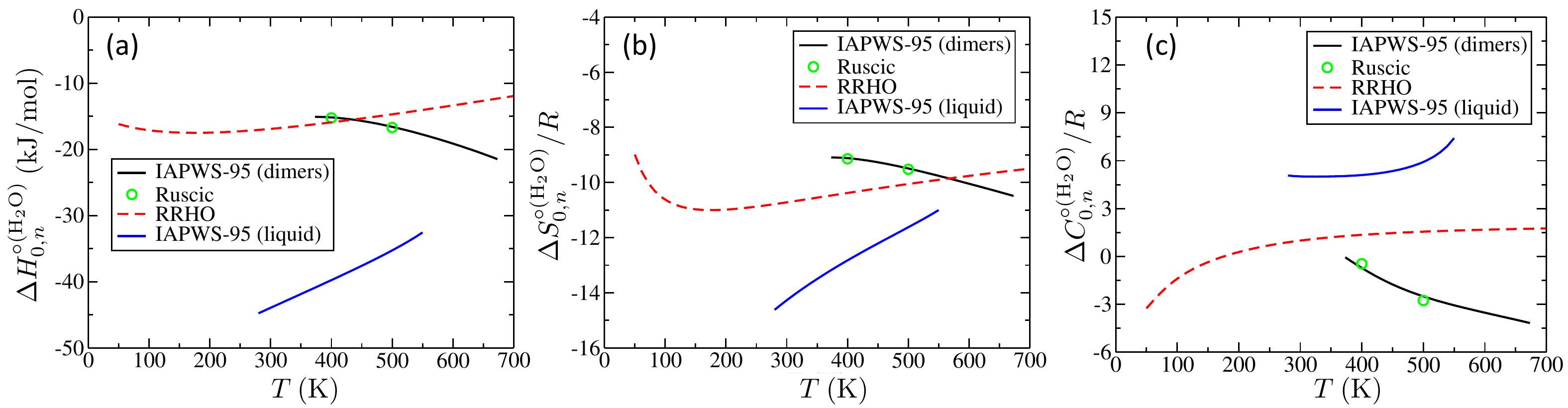}\caption{\label{fig:H2Odim_form}
Standard molar enthalpy, entropy and heat
capacity of hydration of pure-water clusters are plotted as functions
of temperature in panels (a), (b) and (c), respectively. Thermodynamic
parameters for dimer formation, extracted from IAPWS-95 in this
work and adopted from Ref. \citep{Ruscic-2013-11940}, are shown by
the black lines and green circles, respectively. The corresponding
results of quantum chemical calculations are given by the dashed red
lines. The thermodynamics of a single hydration step of a very large
pure-water cluster is represented by the blue lines.}
\end{figure*}

Using Eq. (\ref{eq:nFliq}), the change in the molar Gibbs free energy in the process of adding a water molecule to
a large $(0,n)$-cluster can be calculated as
\begin{equation}
\Delta G_{0,n}^{\circ({\rm H_{2}O})}=G_{0,n}^{\circ}-G_{0,1}^{\circ}-G_{0,n-1}^{\circ}=G_{{\rm liq}}-G_{0,1}^{\circ}.
\end{equation}
Temperature derivatives of this expression produce the enthalpy, entropy and constant-pressure heat capacity changes, $\Delta H_{0,\infty}^{\circ}$, $\Delta S_{0,\infty}^{\circ}$ and $\Delta C_{0,\infty}^{\circ}$ ($n$ is substituted with $\infty$ to emphasize the large size of clusters), plotted in Figure \ref{fig:H2Odim_form} by the blue lines.
Corresponding changes in standard molar enthalpy, entropy and constant-pressure
heat capacity at the reference temperature are shown in the last row
of Table \ref{tab:water_clusters}.
\begin{table*}[!htb]
\centering
\begin{tabular}{|c|c|c|c|}
\hline 
 & $\Delta H_{0,n}^{\stst({\rm H_{2}O)}}$ (kJ/mol) & $\Delta S_{0,n}^{\stst({\rm H_{2}O)}}/R$ & $\Delta C_{0,n}^{\stst({\rm H_{2}O)}}/R$ \tabularnewline
\hline 
\hline 
$n=2$ & $-16.59$ & $-9.49$ & $-2.51$\tabularnewline
\hline 
$n=\infty$ & $-35.26$ & $-11.62$ & $5.92$\tabularnewline
\hline 
\end{tabular}\caption{\label{tab:water_clusters}Standard molar enthalpy, entropy and heat
capacity of hydration of pure-water clusters, evaluated from IAPWS-95
at $T_{{\rm ref}}$.}
\end{table*}
These three parameters can be used in an equation similar to Eq.
(\ref{eq:dG_1n_T}) to approximate the temperature dependence of $\Delta G_{0,\infty}^{\circ({\rm H_{2}O})}$.

It is instructive to attempt to use Eq. (\ref{eq:nFliq}) in order to calculate the pressure of pure water vapor from Eq. (\ref{eq:P_tot}) in the large fugacity regime where the contribution of large water clusters to the total pressure is expected to be dominant. To this end, we substitute Eqs. (\ref{eq:dG_mn}) and (\ref{eq:nFliq}) into the first r.h.s term of Eq.  (\ref{eq:P_tot}) to obtain
\begin{align}
P_{\rm H_2O}=&P^{\circ}\sum_{n=1}^{\infty}e^{-\Delta G_{0,n}^{\circ}/RT}\left(\frac{f}{P^{\circ}}\right)^{n}\nonumber\\
\approx & P^{\circ}\sum_{n=1}^{\infty}e^{-n(G_{\rm liq}-G_{0,1}^{\circ})/RT}\left(\frac{f}{P^{\circ}}\right)^{n}.
\label{eq:P_H2O_sum}
\end{align}
This constitutes an infinite geometric series which can be summed up exactly to yield, with the help of Eq. (\ref{eq:f_H2O})
\begin{equation}
P_{\rm H_2O}=P^{\circ}\left[e^{(G_{\rm liq}-G_{0,1})/RT}-1\right]^{-1}. \label{eq:P_H2O_summed}
\end{equation}
The pressure of water vapor is thus seen to monotonically increase when the partial molar Gibbs free energy of ideal water vapor approaches that of liquid water from below. The pressure diverges, indicating the limit of applicability of the virial expansion in Eq. (\ref{eq:P_H2O_sum}), at $G_{0,1}=G_{\rm liq}$, which is the vapor-liquid phase boundary.

\subsection{Water dimer formation}

Now consider the formation of the smallest possible pure-water
clusters, i.e., dimers. As discussed above, the deviation
of the Gibbs free energy for water from that of ideal vapor (black and dashed blue lines, respectively, in Figure \ref{fig:water_GSCp}(a)), at pressures just below the boiling point ($P\apprle28\,{\rm bar}$)
is due to the non-ideality of water vapor. The pressure
of non-ideal vapor can be approximated by truncating the summation
in the first r.h.s term of Eq. (\ref{eq:P_tot}) to only monomers
and dimers, resulting in $P\approx f+e^{-\Delta G_{0,2}^{\circ}/RT}f^{2}/P^{\circ}$.
Here, the fugacity $f$ of the non-ideal vapor is exactly the pressure
of the ideal vapor at the same Gibbs free energy, and so one can extract
$\Delta G_{0,2}^{\circ}=\Delta G_{0,2}^{\circ({\rm H_{2}O})}$ by
evaluating the low-pressure limit of $(P_{{\rm steam}}-f)/f^{2}$
from the difference between the solid black and dashed blue lines
in Figure \ref{fig:water_GSCp}(a). The low-pressure limit is required
to eliminate the contribution of trimers and larger clusters. Finding
$\Delta G_{0,2}^{\circ({\rm H_{2}O})}$ as a function of temperature
and numerically evaluating its temperature derivatives provides the changes
in other thermodynamic parameters for the same process. Standard molar
enthalpy, entropy and heat capacity changes, extracted from IAPWS-95
using this approach, are shown in Figure \ref{fig:H2Odim_form} by the
black lines. Standard molar enthalpy, entropy and heat capacity changes
evaluated at $T_{{\rm ref}}$ are given in the upper numerical row
of Table \ref{tab:water_clusters}. The temperature dependence of heat
capacity for a water molecule in the dimer is depicted in Figure \ref{fig:water_CvCp}
by the red line. Essentially the same approach, also based on IAPWS-95,
was used in Ref. \citep{Ruscic-2013-11940}. Two representative temperature
points from this reference are shown in Figure \ref{fig:H2Odim_form}
by green circles.

To have an independent source of thermodynamic parameters
for the cluster formation processes, we performed quantum chemical
calculations using the Gaussian 16 software package \citep{Gaussian16}. The computational
method used was Density Functional Theory \citep{Parr-1989-Density},
performed with the B3LYP \citep{Becke-1993-5648,Lee-1988-785} functional
and a 6-311+G(d,p) basis set for all atoms. Structural optimizations
and vibrational calculations were all performed at this level. The
grid used for numerical integration of exchange-correlation energies
was ultrafine, and the spatial step-size during optimization was held
as small as possible. Chloride complexes and water clusters were explicitly
hydrated by building up structures one water molecule at a time. For
example, the lowest energy structure for ${\rm X}\text{:}({\rm H_{2}O})_{n}$
was found by placing an explicit water molecule at several orientations
(5 -- 15) around the ${\rm X\text{:}(H_{2}O)}_{n-1}$ cluster, and
allowing each of these initial guesses to optimize independently.
The ${\rm X}\text{:}({\rm H_{2}O})_{n}$ structure with the lowest
``cold'' energy (energy at $T=0$ with zero-point energies of vibrations
excluded) was then chosen, and used to construct the initial guesses
for ${\rm X\text{:}(H_{2}O)}_{n+1}$. This approach samples many possible
geometries and maximizes the chances of finding the true global minimum
for a given cluster. All optimizations were confirmed to be at an
energy minima by a subsequent vibrational calculation that resulted
in no imaginary frequencies.

The total ground-state energy of a cluster in its optimal geometry,
the moment of inertia tensor and the vibrational frequencies were
extracted from Gaussian 16 output files. These parameters were then
used to generate a complete internally consistent EOS for an ideal
gas of clusters of a given size within the rigid rotor harmonic oscillator
(RRHO) approximation \citep{Ochterski-2000-Thermochemistry} using the code Magpie \citep{Ticknor-2020-proc}.
The ground electronic state was always assumed. Such thermochemical
calculations are illustrated in Figure \ref{fig:H2Odim_form}, where
the standard molar enthalpy, entropy and heat capacity changes, corresponding
to the formation of a water dimer, Eq. (\ref{eq:H2Odim_form}), are
shown by the dashed red lines in panels (a), (b) and (c), respectively. As one can see, the magnitudes of enthalpy and entropy changes for the dimer formation process, obtained using the two described methods,
agree reasonably well. Their temperature dependences and therefore the change
in heat capacity are rather different however. In particular, $\Delta C_{0,2}^{\circ({\rm H_{2}O})}$
is of different sign for RRHO and IAPWS-95 based approaches. This
disagreement is likely originating from the well-known failure of
RRHO to account for the anharmonicity of the low-frequency inter-molecular
vibrations \citep{Ruscic-2013-11940}. 

\section{Thermodynamics of ${\rm X}_{ m}$ cluster formation\label{sec:X_therm}}

\begin{figure*}[!htb]
\centering
\includegraphics[width=0.8\paperwidth]{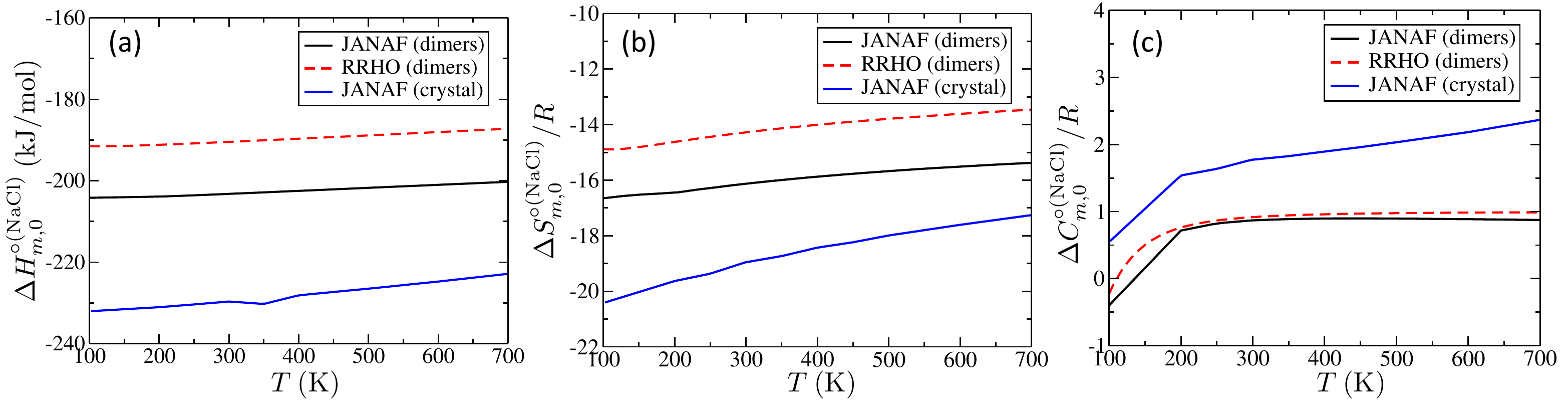}\caption{\label{fig:NaCldimer_form}
$\Delta H_{m,0}^{\circ({\rm NaCl})}$, $\Delta S_{m,0}^{\circ({\rm NaCl})}$ and $\Delta C_{m,0}^{\circ({\rm NaCl})}$
for the process of addition of a ${\rm NaCl}$ monomer to a pure ${\rm NaCl}$
cluster, Eq. (\ref{eq:react_salt}) at $n=0$. Thermodynamic state
functions are extracted from JANAF tables for $m=2$ (black lines),
and $m=\infty$ (blue lines). RRHO results for the dimer formation
($m=2$) are represented by the dashed red lines.}
\end{figure*}

Now consider dimerization of ${\rm NaCl}$ molecules in an anhydrous
vapor of sodium chloride. RRHO results for this process are shown by dashed red lines
in Figure \ref{fig:NaCldimer_form}.
Thermochemical data for ${\rm NaCl}$ monomers, dimers and crystalline
bulk are available from the JANAF thermochemical tables \citep{JANAF-1998}.
This allows one to evaluate the thermodynamics of formation of dimers
and very large clusters out of the ideal gas of monomers. The former, plotted by solid black lines in Figure \ref{fig:NaCldimer_form}, shows the reasonable agreement with the RRHO results for the enthalpy
and entropy, and good agreement for the heat capacity.

The thermodynamics for the process of adding a ${\rm NaCl}$
monomer to a large cluster, Eq. (\ref{eq:react_salt}) with $m\gg1$
and $n=0$, is shown by the blue lines in the same figure. The caveat
is that since the thermodynamic data for ${\rm NaCl}$ is only available
at standard pressure from the tables, one cannot extrapolate to zero
pressure as we did for large pure-water clusters. However, the compressibility
of crystalline ${\rm NaCl}$ is lower than that of liquid water by at least
an order of magnitude so we simply assume that the bulk ${\rm NaCl}$
thermodynamics potentials are the same at zero and standard pressures.
It is expected that an added monomer interacts with more salt molecules
in a larger cluster than in a dimer, which rationalizes the observation
of $\Delta H_{\infty,0}^{\circ({\rm NaCl})}<\Delta H_{2,0}^{\circ({\rm NaCl})}$
and $\Delta C_{\infty,0}^{\circ({\rm NaCl})}>\Delta C_{2,0}^{\circ({\rm NaCl})}$
in the figure. On the other hand, the motion of salt molecules in
the large cluster is more correlated than in the dimer due to the
stronger interaction in the former case, which is the reason for $\Delta S_{\infty,0}^{\circ({\rm NaCl})}<\Delta S_{2,0}^{\circ({\rm NaCl})}$
in Figure \ref{fig:NaCldimer_form}(b).

The knowledge of the dimer formation thermodynamics allows one to estimate the contribution of $(2,n)$-clusters to $P_{{\rm X}}$. Eq. (\ref{eq:P_nm}) yields $P_{2,n}/P^{\circ}=e^{-\Delta\tilde{G}_{2,n}^{\circ}/RT}\left(\frac{f}{P^{\circ}}\right)^{n}$,
where
\begin{align}
\Delta\tilde{G}_{2,n}^{\circ}&=2\Delta G_{s}^{\circ}+\Delta G_{2,0}^{\circ({\rm NaCl})}+\Delta G_{2,1}^{\circ({\rm H_{2}O})}\nonumber\\
&+\Delta G_{2,2}^{\circ({\rm H_{2}O})}+...+\Delta G_{2,n}^{\circ({\rm H_{2}O})}.\label{eq:dG_2n}
\end{align}
The last $n$ r.h.s. terms in this expression describe the successive
addition of water molecules to the ${\rm NaCl}$ dimer, stabilizing
the latter. The stabilization effect is expected to be greater for
a single ${\rm NaCl}$ molecule than for the dimer because in the
latter case the interaction between a salt molecule and water is screened by the other salt molecule. We can therefore estimate $P_{2,n}$
from above by substituting $\Delta G_{2,i}^{\circ({\rm H_{2}O})}$
with $\Delta G_{1,i}^{\circ({\rm H_{2}O})}$ , which transforms Eq.
(\ref{eq:dG_2n}) into
\begin{equation}
\Delta\tilde{G}_{2,n}^{\circ}\approx\Delta G_{s}^{\circ}+\Delta G_{2,0}^{\circ({\rm NaCl})}+\Delta\tilde{G}_{1,n}^{\circ}
\end{equation}
which in turn produces
\begin{equation}
P_{2,n}/P^{\circ}\approx e^{-\left[\Delta G_{s}^{\circ}+\Delta G_{2,0}^{\circ({\rm NaCl})}\right]/RT}P_{1,n}/P^{\circ}.\label{eq:P_2n}
\end{equation}
We estimate the value of the exponential function in the r.h.s. of Eq. (\ref{eq:P_2n})
at $T_{{\rm ref}}$ using the data from Figure \ref{fig:NaCldimer_form}
and Table \ref{tab:sublimation} to be $\approx 0.01$ so $P_{2,n}$
is smaller than $0.01P_{1.n}$. Therefore, the ${\rm NaCl}$ solubility
in water vapor is strongly dominated by $(1,n)$-clusters. We expect
this statement to be accurate at other temperatures, as well as for
${\rm CuCl}$. The one caveat in the above considerations is that
it is assumed that two salt monomers are bonded inside a water cluster.
However, if there is large water cluster formed (effectively a droplet
of liquid water), there is a gain in translational entropy if the
salt dimer is split into two monomers that can explore the volume
of the droplet independently. Furthermore, dissociation of neutral
salt monomers into atomic ions can be expected in such droplets. This
is expected to become important only very close to the vapor-liquid
phase transition, i.e., for highly non-ideal vapor containing a lot
of droplets. We do not consider this regime here.

\begin{figure}[!htb]
\centering
\includegraphics[width=0.38\paperwidth]{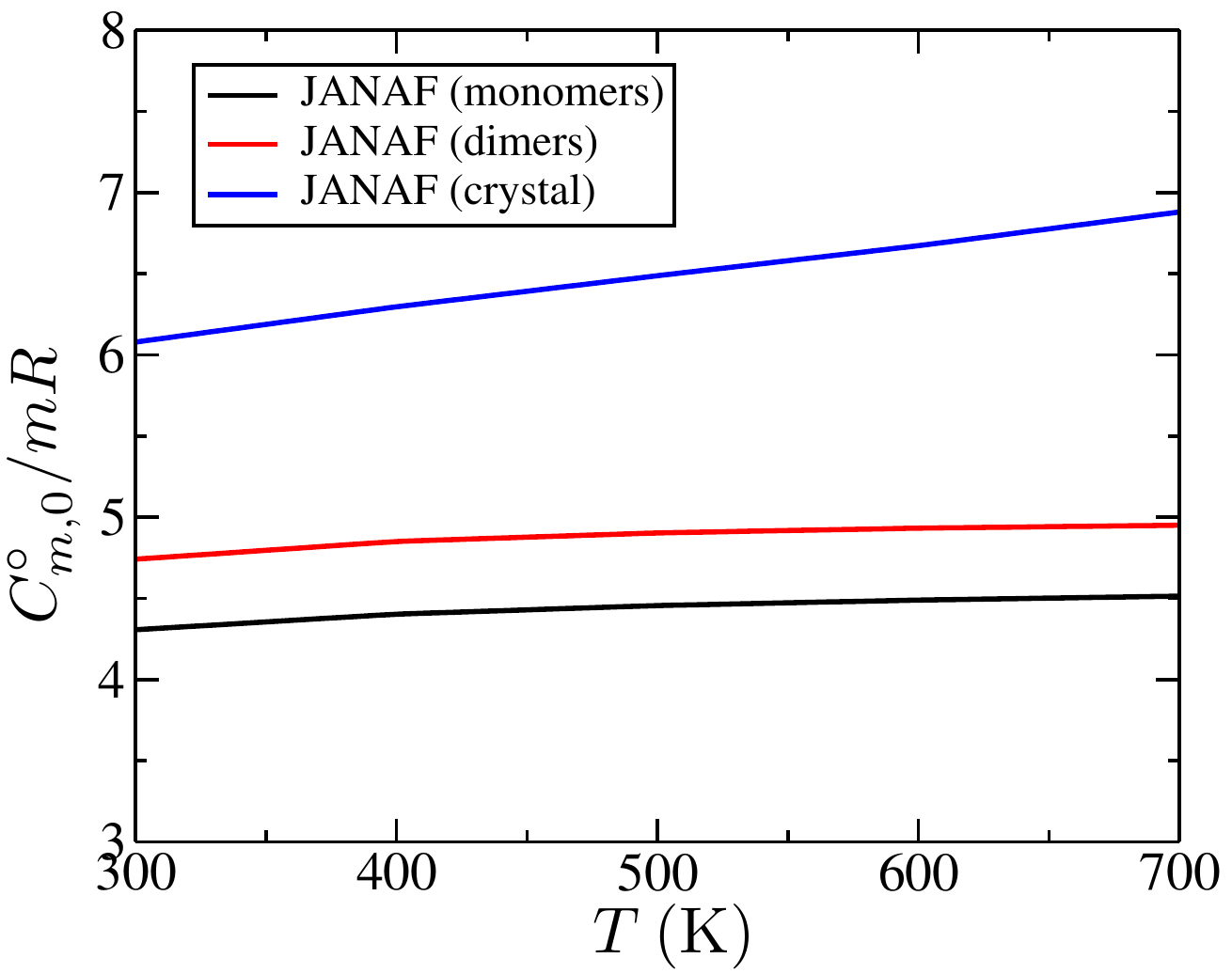}\caption{\label{fig:NaCl_Cp}Standard molar heat capacity of a pure-${\rm NaCl}$
cluster normalized per number of salt monomers in it, $C_{m,0}^{\circ}/mR$.
The results for monomers ($m=1$), dimers ($m=2$) and crystalline
phase ($m\rightarrow\infty$) are plotted by the black, red and blue
lines, respectively.}

\end{figure}

Assuming classical harmonic intra- and inter-molecular vibrations,
the molar heat capacity of an ideal gas of ${\rm NaCl}$ monomers
is $C_{1,0}^{\circ}=\frac{3}{2}R+\frac{2}{2}R+\frac{1}{1}R+R$, where
the r.h.s. terms stand for, respectively, the contributions of three
translational degrees of freedom (DOF), the rotation of a linear rotor,
the single vibrational mode, and the difference between constant-volume
and constant-pressure heat capacities of an ideal gas. This results
in $C_{1,0}^{\circ}/R=4.5$. For a dimer, and larger clusters, similar
considerations yield (assuming a non-linear rotor) $C_{m,0}^{\circ}=\frac{3}{2}R+\frac{3}{2}R+\frac{6m-6}{1}R+R=(6m-2)R$,
which results in $C_{2,0}/2R=5$ and $C_{m,0}/mR=6R$ for $m\rightarrow\infty$.
These simple estimates are seen to agree well with results calculated
from the JANAF tables \citep{JANAF-1998} in Figure \ref{fig:NaCl_Cp}.
Some deviations are due to anharmonicity in large clusters and partially
quantum-mechanically ``frozen'' vibrational modes in small clusters.
The changes in heat capacities as $\Delta C_{m,0}^{\circ({\rm NaCl})}=C_{m,0}^{\circ}-C_{m-1,0}^{\circ}-C_{1,0}^{\circ({\rm NaCl})}$
result in the sold black and blue lines in Figure \ref{fig:NaCldimer_form}(c).
The agreement of the RRHO results for the dimer with the black line
is also good, implying that the assumption of harmonic vibrations in
RRHO is accurate in small clusters.

An application of the same DOF-counting approach to evaluation of
heat capacities of pure-water clusters produces
\begin{equation}
C_{0,n}^{\circ}=\frac{3}{2}R+\frac{3}{2}R+\frac{6n-6}{1}R+R,
\end{equation}
where the inter-molecular vibrations ($6n-6$ DOFs) are all assumed
classical and harmonic, and the intra-molecular vibrations are all
assumed quantum-mechanically frozen and, therefore, not contributing
to the heat capacity. The results are $C_{0,1}^{\circ}/R=4$, $C_{0,2}^{\circ}/2R=5$
and $C_{0,n}^{\circ}/nR=6$ for $n\rightarrow\infty$. These could
be compared to the results extracted from IAPWS-95, Figure \ref{fig:water_CvCp}.
The agreement is good for the ideal gas of monomers. The agreement
is worse for the dimers and, especially, for the very large clusters.
The difference is expected to be due to significant anharmonicity
of van der Waals interaction between water molecules in clusters.
The same anharmonicity is the chief reason for disagreement between
IAPWS-95 and RRHO results for dimers in Figure \ref{fig:H2Odim_form}.

The thermodynamics of successive addition of water molecules to a
salt-bearing water cluster ${\rm X}\text{:}({\rm H_{2}O})_{n}$ should converge to that of pure-water clusters at large $n$. The above
considerations of heat capacities demonstrated that the thermodynamics
of water clusters is complex so, for example, simple estimates of
the cluster heat capacities based on counting DOFs are inaccurate.
We postulate that the agreement between such simple estimates and
experimental data for heat capacities of ${\rm X}\text{:}({\rm H_{2}O})_{n}$
in Ref. \citep{Pitzer-1986-1445} are due to error cancellation.

\subsection{Thermodynamics of sublimation\label{subsec:Sublimation}}

\begin{figure}[!htb]
\centering
\includegraphics[width=0.38\paperwidth]{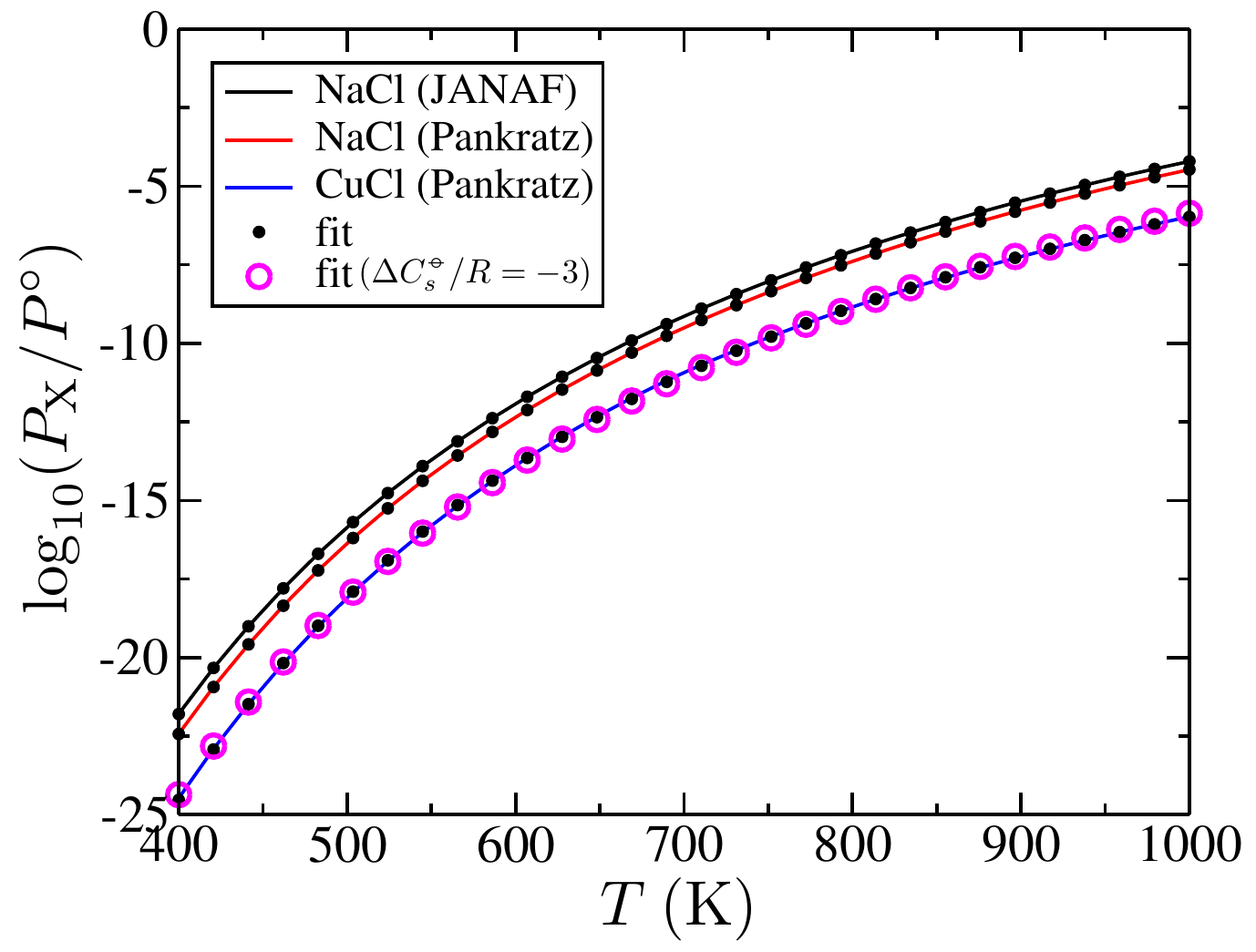}\caption{\label{fig:subl_fit}Pressure of anhydrous ${\rm NaCl}$ and ${\rm CuCl}$
monomer vapors in equilibrium with crystalline ${\rm NaCl}$ and ${\rm CuCl}$,
respectively, as a function of temperature. The data obtained from
the JANAF \citep{JANAF-1998} and Pankratz \citep{Panktratz-1984-Halides}
thermochemical tables are depicted by solid lines. Black dots represent
the fitting results where the enthalpy, entropy and heat capacity
of sublimation were all allowed to vary. Magenta circles depict the
fitting result where only the enthalpy and entropy of sublimation
were allowed to vary, but the heat capacity of sublimation was set
to $\Delta C_{s}^{\stst}/R=-3$.}
\end{figure}

Thermodynamics of sublimation for ${\rm NaCl}$ and ${\rm CuCl}$
is known accurately from the JANAF \citep{JANAF-1998} and Pankratz \citep{Panktratz-1984-Halides}
thermochemical tables. In particular, $\Delta G_s^\circ(T)$ can be extracted from the tables over a wide temperature range. However, (i) to treat all the thermochemical data on the
same level of approximations, and also (ii) to anticipate situations
where such detailed thermochemical data on temperature variation is
not available, we will use an expression similar to Eq. (\ref{eq:dG_1n_T}) to represent $\Delta G_{s}^{\circ}(T)$
instead of the full thermochemical data.
To properly parameterize such an expression, we first generate $P_{{\rm X}}(T)$
from Eq. (\ref{eq:Px}) at vanishing water fugacity over an experimentally
realistic temperature range using the full thermochemical data from the JANAF \citep{JANAF-1998} and Pankratz \citep{Panktratz-1984-Halides} tables. Then, the resulting $P_{{\rm X}}(T)$ is fit with the combination
of Eq. (\ref{eq:Px}) and Eq. (\ref{eq:dG_1n_T}). The fitting procedure
will be described in detail in Sec. \ref{subsec:fitting}. The results
of the fitting are compiled in Table \ref{tab:sublimation}, and the
fit qualities can be seen in Figure \ref{fig:subl_fit}.
\begin{table*}[!htb]
\centering
\begin{tabular}{|c|c|c|c|c|c|}
\hline 
 & ${\rm NaCl}$ (J) & ${\rm NaCl}$ (P) & ${\rm CuCl}$ (J) & ${\rm CuCl}$ (P) & ${\rm CuCl}$ (P)\tabularnewline
\hline 
\hline 
$\Delta H_{s}^{\stst}\,{\rm (kJ/mol)}$  & $227$ & $231$ & $225$ & $243$ & $239$\tabularnewline
\hline 
$\Delta S_{s}^{\stst}/R$ & $18.0$ & $18.0$ & $16.9$ & $17.0$ & $15.5$\tabularnewline
\hline 
$\Delta C_{s}^{\stst}/R$ & $-2.28$ & $-2.31$ & $-2.97$ & $-7.59$ & $-3$ (forced)\tabularnewline
\hline 
\end{tabular}\caption{\label{tab:sublimation}Standard molar enthalpy, entropy and heat
capacity of sublimation, Eq. (\ref{eq:react_sublim}), obtained from
fitting the solid lines in Figure \ref{fig:subl_fit}. (J) and (P) stand
for the JANAF \citep{JANAF-1998} and Pankratz \citep{Panktratz-1984-Halides}
thermochemical tables. The last column corresponds to fitting where
the heat capacity of sublimation was forced to be $\Delta C_{s}^{\stst}/R=-3$
when fitting.}
\end{table*}
Pressure evaluated directly from the thermochemical tables is shown
in Figure \ref{fig:subl_fit} by solid lines. The fitting results are
shown by black dots and magenta circles. The agreement is excellent for both ${\rm NaCl}$ and ${\rm CuCl}$, even though Eq. (\ref{eq:dG_1n_T}) is a smooth function of temperature, and, therefore, might not {\it a priori} be considered applicable to describe the thermodynamics of ${\rm CuCl}$, which has a solid-to-liquid transition at $T\approx 700\:{\rm K}$.
In Table \ref{tab:sublimation}, the first four numerical columns represent the fitting
of ${\rm NaCl}$ and ${\rm CuCl}$ sublimation thermodynamics, with
thermochemical data coming Refs. \citep{Panktratz-1984-Halides,JANAF-1998},
where all the three parameters $\Delta H_{s}^{\stst}$, $\Delta S_{s}^{\stst}$
and $\Delta C_{s}^{\stst}$ are allowed to vary. Theses results are
shown by black dots in the Figure.
As is seen in Table \ref{tab:sublimation}, fitting the Pankratz thermochemical  data for ${\rm CuCl}$ results in a significantly larger magnitude of the change in the heat capacity upon sublimation (fourth numerical column), compared to those in the first three numerical columns.
To test the sensitivity of the fitting quality
to this parameter, we also fit the same thermochemical data with the constraint $\Delta C_{s}^{\stst}/R=-3$.
The result of this fit, the last column in Table \ref{tab:sublimation},
is represented by magenta circles in Figure \ref{fig:subl_fit}. The
deviation from the direct thermochemical data, and from the fit with
all the three parameter varied, is almost negligible, indicating that the magnitude of
$\Delta C_{s}^{\stst}$ extracted from this fitting is not too reliable. 

The condensation of gaseous ${\rm NaCl}$ into
a crystal (i.e., reverse sublimation) can be thought of as adding
a single salt molecule to a very large ${\rm NaCl}$ cluster, i.e.,
we expect $\Delta G_{s}^{\circ}=-\Delta G_{\infty,0}^{\circ({\rm NaCl})}$,
and similarly for $\Delta H^{\circ}$ and $\Delta S^{\circ}$. It
is thus instructive to compare the thermodynamics of reverse sublimation
(values in Table \ref{tab:sublimation} taken with the opposite sign)
to that of formation of ${\rm NaCl}$ dimer. In particular, salt molecules
in a dimer are expected to interact more loosely than in bulk, which
should result in $\Delta H_{2,0}^{\circ({\rm NaCl})}>-\Delta H_{s}^{\circ}$,
and $\Delta S_{2,0}^{\circ({\rm NaCl})}>-\Delta S_{s}^{\circ}$. Comparing
Figure \ref{fig:NaCldimer_form} and Table \ref{tab:sublimation} we
see that $\Delta H^{\circ({\rm NaCl})}$ and $\Delta S^{\circ({\rm NaCl})}$
are indeed larger for the dimer than for bulk, although not by much. 

\section{Solubility of Sodium Chloride in Water Vapor\label{sec:NaCl_solubility}}

\begin{figure*}[!htb]
\centering
\includegraphics[width=0.75\paperwidth]{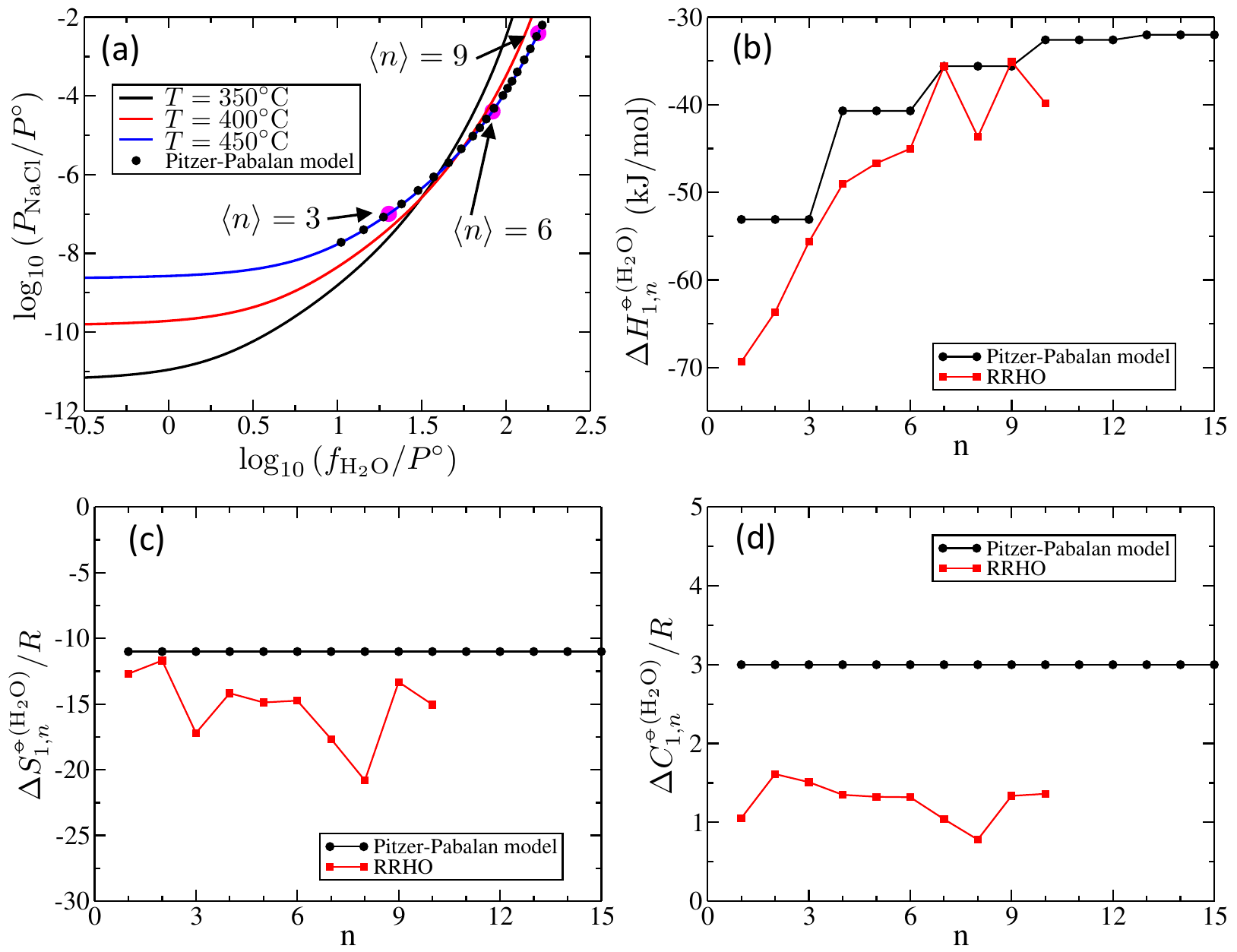}\caption{\label{fig:synthetic_PP86}
(a) Partial pressure of ${\rm NaCl}$-bearing water clusters generated using the parameters from the last three columns in Table \ref{tab:PP86_parms} (solid lines). Black dots represent the data directly digitized from Figure 2 in Ref. \citep{Pitzer-1986-1445}. Thermodynamic parameters from
second, third and forth numerical columns in Table \ref{tab:PP86_parms}
are compared to the RRHO results in panels (b), (c) and (d), respectively.}
\end{figure*}

In their seminal paper \citep{Pitzer-1986-1445}, Pitzer and Pabalan
were able to reproduce various experimental data for ${\rm NaCl}$
solubility in water vapor using the imperfect gas theory outlined above.
The enthalpy, entropy and heat capacity changes, obtained for successive addition of water molecules to a cluster, are given
in the first, third and fourth numerical columns, respectively, of
Table \ref{tab:PP86_parms}.
\begin{table*}[!htb]
\centering
\begin{tabular}{|c|c|c|c|c|}
\hline 
 & $\Delta H_{1,n}^{\stst({\rm H_{2}O})}$ (kJ/mol) & $\Delta H_{1,n}^{\stst({\rm H_{2}O})}$ (kJ/mol), corrected & $\Delta S_{1,n}^{\stst({\rm H_{2}O)}}/R$ & $\Delta C_{1,n}^{\stst({\rm H_{2}O)}}/R$\tabularnewline
\hline 
\hline 
$n=1\text{ - }3$ & $-53.1$  & $-53.1$  & \multirow{4}{*}{-11} & \multirow{4}{*}{3}\tabularnewline
\cline{1-3} 
$n=4\text{ - }6$ & $-40.7$ & $-40.7$  &  & \tabularnewline
\cline{1-3} 
$n=7\text{ - }9$ & $-35.8$  & $-35.6$  &  & \\
\cline{1-3} 
$n=10\text{ - }\infty$ & $-32.8+0.42\times{\rm floor}\left(\frac{n-10}{3}\right)$ & $-32.6+0.58\times{\rm floor}\left(\frac{n-10}{3}\right)$ &  & \rule{0pt}{2.6ex}\rule[-1.2ex]{0pt}{0pt} \\
\hline 
\end{tabular}\caption{\label{tab:PP86_parms}Thermodynamic parameters of hydration of salt-bearing
water clusters for the original Pitzer-Pabalan model \citep{Pitzer-1986-1445}.
First numerical column is original enthalpies from Ref. \citep{Pitzer-1986-1445}.
The floor function is denoted by ${\rm floor}()$. Second
numerical column is the corrected version of the first column. The
last two columns are $n$-independent standard molar entropy and heat
capacity of hydration.}
\end{table*}
In order to reproduce the Pitzer-Pabalan modeling results we needed
to slightly modify some of their originally reported enthalpies. We
suspect that the values were rounded off too coarsely in Table 2 of
the original publication \citep{Pitzer-1986-1445}. The new corrected
enthalpies are given in the second numerical column of Table \ref{tab:PP86_parms}.
The parameters from the second, third and fourth numerical columns
of the table are plotted by black dots in panels (b), (c) and (d),
respectively, of Figure \ref{fig:synthetic_PP86}. Using these corrected
values (entropy and heat capacity values are left the same), along
with those in the first column of Table \ref{tab:sublimation} to calculate
$\Delta G_{s}^{\circ}(T)$, we reproduced Pitzer and Pabalan's results
for the partial pressure of ${\rm NaCl}$-bearing water clusters.
In particular, the dependence of $P_{{\rm NaCl}}$ on water fugacity,
Eq. (\ref{eq:Px}), for three different temperatures is depicted in
Figure \ref{fig:synthetic_PP86}(a) by the solid lines. Results at $T=450^{\circ}{\rm C}$
from Figure 2 in Ref. \citep{Pitzer-1986-1445} are plotted as
black dotes in Figure \ref{fig:synthetic_PP86}(a). As can be seen,
it agrees perfectly with what is calculated in this work using the
corrected values of thermodynamic parameters (blue line).
To explore questions related to possible overfitting, $P_{{\rm NaCl}}$ dependences on $f$ for a
range of temperatures were generated from the Pitzer-Pabalan model using
the parameters given in the last three columns of Table \ref{tab:PP86_parms}.
Those data, treated as if they were experimental results, were then used to test the model developed in
this work. This approach allows for a direct comparison
with the original Pitzer-Pabalan results, see e.g., Figure \ref{fig:NaCl_fit_kinks}(b)
below. 

The dependence of $P_{{\rm NaCl}}$ on temperature (at fixed $f$) is
qualitative different at lower and higher fugacities as seen in Figure \ref{fig:synthetic_PP86}.
The temperature derivative of Eq. (\ref{fig:synthetic_PP86}) is
\begin{equation}
\left(\frac{\partial P_{{\rm X}}}{\partial T}\right)_{f}=P^{\circ}\sum_{n=0}^{\infty}\frac{\Delta\tilde{H}_{1,n}^{\circ}}{RT^{2}}e^{-\Delta\tilde{G}_{1,n}^{\circ}/RT}\left(\frac{f}{P^{\circ}}\right)^{n},\label{eq:dPx_dT}
\end{equation}
where, as defined above, $\Delta\tilde{H}_{1,n}^{\circ}=\Delta H_{s}^{\circ}+\Delta H_{1,1}^{\circ({\rm H_{2}O})}+\Delta H_{1,2}^{\circ({\rm H_{2}O})}+...+\Delta H_{1,n}^{\circ({\rm H_{2}O})}$.
For clusters with a small number of water molecules, this sum is dominated
by a large positive $\Delta H_{s}^{\circ}$, yielding $\Delta\tilde{H}_{1,n}^{\circ}>0$.
On the other hand, $\Delta H_{1,n}^{\circ({\rm H_{2}O})}$ are negative due to the exothermicity of hydration,
so $\Delta\tilde{H}_{1,n}^{\circ}$ becomes negative for larger clusters.
Mean cluster size, defined by Eq. (\ref{eq:mean_n}),
increases with fugacity, so it is expected that at high fugacities
the summation in Eq. (\ref{eq:dPx_dT}) is dominated by large clusters
with negative $\Delta\tilde{H}_{1,n}^{\circ}$, thus resulting in
negative $\left(\frac{\partial P_{{\rm X}}}{\partial T}\right)_{f}$
and so retrograde solubility is observed. For small clusters,
however, the temperature dependence of $P_{{\rm X}}$ is dominated
by the thermodynamics of sublimation (Figure \ref{fig:subl_fit}), and
so $P_{{\rm X}}$ grows rapidly with temperature at low fugacities
and the solubility is ``normal'' (i.e., not retrograde). Magenta
circles in Figure \ref{fig:synthetic_PP86}(a) denote fugacities where
$\langle n\rangle=3,6,9$ at $T=450\,{\rm C^{\circ}}$. The temperature
derivative of $P_{{\rm X}}$ is seen to change sign at $\langle n\rangle\sim4-5$.
In agreement with this, comparison of data in Table \ref{tab:sublimation}
and Figure \ref{fig:synthetic_PP86}(b) implies that $\Delta\tilde{H}_{1,n}^{\stst}$ is negative for $n\gtrsim5$.
As $f\rightarrow 0$, Eq. (\ref{eq:Px}) reduces to fugacity-independent
$P_{{\rm X}}=P^{\circ}e^{-\Delta G_{s}^{\circ}/RT}$, which is the
pressure of the anhydrous salt vapor. Accordingly, $P_{{\rm NaCl}}$
becomes constant at low fugacities in Figure \ref{fig:synthetic_PP86}(a).
One can verify that the values of $P_{{\rm NaCl}}$ at $f\rightarrow0$,
taken as a function of temperature, coincide with the black line in
Figure \ref{fig:subl_fit}.

As described above, we employed quantum chemistry calculations (Gaussian
16) together with the thermochemical code Magpie to calculate
the thermodynamics of ${\rm NaCl}$-bearing water clusters within
the RRHO approximation. The results of these calculations, plotted
by red squares in Figure \ref{fig:synthetic_PP86}(b)-(d), are seen
to be in reasonable agreement with the Pitzer-Pabalan model. Notable
deviations for the entropy and for heat capacity
could be rationalized as follows. A known deficiency of the RRHO approach is that it approximates all the vibrational
modes as strictly harmonic \citep{Ochterski-2000-Thermochemistry,Ruscic-2013-11940}.
However, soft vibrational modes of van der Walls-bonded clusters are
expected to be rather anharmonic (e.g., internal hindered rotations),
and so RRHO is expected to underestimate the change in entropy and
heat capacity of a cluster upon addition of a water molecule. This
is in qualitative agreement with what is observed in Figure \ref{fig:synthetic_PP86}(c)
and (d). Another source of inaccuracy of the RRHO calculations in
this work is that, as discussed above, it takes into account only a
single cluster of all the clusters with the same stoichiometry - the
one with the lowest energy. Not including any energetically sub-optimal
clusters (i.e., via the Boltzmann distribution of all accessible configurations
for the same stoichiometry) is expected to result in underestimating
the enthalpy and entropy of hydration, which is again in qualitative agreement with Figure \ref{fig:synthetic_PP86}(c)
and (d).

General considerations for the dependence of $\Delta H_{1,n}^{\stst({\rm H_{2}O})}$,
$\Delta S_{1,n}^{\stst({\rm H_{2}O})}$ and $\Delta C_{1,n}^{\stst({\rm H_{2}O})}$
on the number of water molecules $n$ are as follows. The enthalpy
of hydration $\Delta H_{1,n}^{\stst({\rm H_{2}O})}$ has to be negative
due to attractive interaction of a cluster with an added water molecule. Due to Coulomb screening
of interaction with the salt molecule, this interaction is weaker
for larger clusters so the magnitude of $\Delta H_{1,n}^{\stst({\rm H_{2}O})}$
decreases with $n$. The entropy change $\Delta S_{1,n}^{\stst({\rm H_{2}O})}$
is expected to be strongly negative since the large translational
entropy of a water molecule is substituted with the much smaller vibrational
one. The standard molar translational entropy of an ideal gas of water
molecules at the reference temperature $T_{{\rm ref}}=500{\rm \,K}$
is evaluated using the Sackur-Tetrode equation \citep{Atkins-2006-Physical}
to be $S_{{\rm 0,1}}^{\stst}/R\approx18.7$. If this entropy is lost
upon addition of the water molecule to a cluster and nothing else
changed, the result would be $\Delta S_{1,n}^{\stst({\rm H_{2}O})}/R=-18.7$,
which is not too inconsistent with the RRHO results in Figure \ref{fig:synthetic_PP86}(c).
The entropy change, extracted by Pitzer and Pabalan, is not as negative, implying that the translational entropy
is not completely lost but rather substituted with a lower, but still
substantial, entropy of soft vibrational modes. Assuming that the three translational and three rotational
degrees of freedom (DOF) of an isolated water molecule become six
classical harmonic vibrational DOFs upon bonding to a cluster, the
heat capacity change would be $\Delta C_{1,n}^{\stst({\rm H_{2}O})}/R=2$,
which is comparable to what is shown in Figure \ref{fig:synthetic_PP86}(d).

It is important to remember that the general considerations just above only concern the general "smooth" dependence of $\Delta H_{1,n}^{\stst({\rm H_{2}O})}$, $\Delta S_{1,n}^{\stst({\rm H_{2}O})}$ and $\Delta C_{1,n}^{\stst({\rm H_{2}O})}$ on $n$. They cannot account for a specific way a water molecule forms bonds with a given cluster, and how this cluster geometry rearranges upon this hydration. It is therefore expected that some ``noise'', originating from the specific details of interaction of water and salt molecules within a cluster,  should be present on top of the general smooth dependence on $n$ discussed above. This ``noise'' is clearly seen on top of the general smooth dependence on $n$ in the RRHO results (red lines) in Figure \ref{fig:synthetic_PP86}(b)-(d).

\subsection{Fitting Procedure\label{subsec:fitting}}

The dependence of $\Delta H_{1,n}^{\stst({\rm H_{2}O})}$
on $n$ reflects the nature of interaction of a water molecule with
a $(1,n-1)$-cluster. Pitzer and Pabalan used the piecewise constant
functional form with all the steps of the same length of 3. This form
is illustrated by black dots in Figure \ref{fig:synthetic_PP86}. To
explore and accurately control the degree of underfitting/overfitting
we need a more general and flexible functional form. We choose a continuous
piecewise linear function, which if comprising $K$ kinks, is specified
by $K$ kink positions, $K+1$ tangents, and one absolute value -
the value of the function at $n=1$. For instance, a $K$-kink representation
of $\Delta H_{1,n}^{\stst({\rm H_{2}O})}$ as a function of $n$ is
specified by its piecewise-constant derivative
\begin{equation}
\frac{\partial\Delta H_{1,n}^{\stst({\rm H_{2}O})}}{\partial n}=\begin{cases}
h_{1}, & \text{if }n<n_{1}\\
h_{i}, & \text{if }n_{i-1}\leq n<n_{i}\\
h_{K+1}, & \text{if }n>n_{K}
\end{cases}\,\left(i\in[2,K]\right),\label{eq:K-kink_H}
\end{equation}
and by the value of $\Delta H_{1,1}^{\stst({\rm H_{2}O})}$. For example,
a $0$-kink representation, i.e., a simple linear dependence, is defined
by two parameters ( $\Delta H_{1,1}^{\stst({\rm H_{2}O})}$ and $h_{1}$),
whereas four parameters are required to define a 2-kink function (
$\Delta H_{1,1}^{\stst({\rm H_{2}O})}$, $h_{1}$, $h_{2}$, $n_{1}$).
Kink positions $n_{i}$ are assumed continuous numbers, which simplifies
fitting by allowing their continuous variation in standard optimization
routines. 
Similar representations could be introduced for $\Delta S_{1,n}^{\stst({\rm H_{2}O})}$
and $\Delta C_{1,n}^{\stst({\rm H_{2}O})}$. We find, however, that
it is sufficient to represent them by a single $n$-independent value
each. 

To extract the parameters, like those in Eq. (\ref{eq:K-kink_H}),
from fitting, one has to define a function of these parameters that
encodes the deviation of the model from the benchmark experimental
results and then minimize that function. For benchmark data, similar
to that in Figure \ref{fig:synthetic_PP86}(a) where we have $P_{{\rm X}}$
as a function of $f$ at several temperatures, such deviation function
is introduced as follows. We first substitute the water fugacity by a new variable
\begin{equation}
x=\frac{\log_{10}\left[f/f^{(min)}\right]}{\log_{10}\left[f^{(max)}/f^{(min)}\right]},
\end{equation}
where $\left(f^{(min)},f^{(max)}\right)$ covers the full experimentally
accessible range of fugacities. Similarly, for the partial pressure
of salt-bearing water clusters we have
\begin{equation}
y=\frac{\log_{10}\left[P_{{\rm X}}/P_{{\rm X}}^{(min)}\right]}{\log_{10}\left[P_{{\rm X}}^{(max)}/P_{{\rm X}}^{(min)}\right]}.
\end{equation}
Upon these substitutions, Figure \ref{fig:synthetic_PP86}(a) - a standard way of presenting $P_{{\rm X}}$
dependence on water fugacity in the literature - effectively becomes
a linear-linear plot (as opposed to e.g., log-log), the abscissa and ordinate ranging from $0$ to $1$. Then, at fixed temperature
$T=T_{i}$, the deviation of the model data for the partial pressure
of salt-bearing clusters (taken as a function of fugacity $f$), from
the experimental results is calculated as 
\begin{equation}
d_{i}=\int_{x_{i,min}}^{x_{i,max}}dx\,\left[y_{i}^{(mod)}(x)-y_{i}^{(exp)}(x)\right]^{2},
\end{equation}
where $x_{i,min}$ and $x_{i,max}$ are the minimum and maximum values
of $x$ available experimentally at temperature $T_{i}$. The integral
is evaluated by interpolating $y_{i}^{(mod)}$ and $y_{i}^{(exp)}$
to make them continuous functions of $x$. The full deviation function
is then evaluated by summing up all the single-temperature deviations
as
\begin{equation}
D=\frac{1}{2}\log_{10}\left[\frac{1}{N}\sum_{i=1}^{N}d_{i}\right],\label{eq:obj_func}
\end{equation}
where $N$ is the number of temperatures. The rationale behind this
definition of deviation can be illustrated by the following example.
Suppose that the model results for $\log_{10}P_{{\rm X}}/P^{\circ}$
deviate by exactly 1\% of $\log_{10}\left[P_{{\rm X}}^{(max)}/P_{{\rm X}}^{(min)}\right]$
from the experimental data at all water fugacities. That would mean
a deviation of the model curves from  experimental by 1\%
of the vertical range in Figure \ref{fig:synthetic_PP86}(a), treating
it as a linear-linear plot. The deviation of this magnitude is still
be rather noticeable in a figure. This deviation produces $d_{i}=10^{-4}$
and therefore $D=-2$.
On the other hand, the deviation of the model
results, represented by $\log_{10}P_{{\rm X}}/P^{\circ}$, by 0.1\%
of $\log_{10}\left[P_{{\rm X}}^{(max)}/P_{{\rm X}}^{(min)}\right]$
from the experimental data, results in $D=-3$ and is perceived as
a perfect agreement between theory and experiment when plotted in
a figure similar to Figure \ref{fig:synthetic_PP86}(a). Therefore,
the value of $D$ represents the decimal logarithm of the deviation
of the model $\log_{10}P_{{\rm X}}$ from its experimental counterpart,
normalized to the entire span of $\log_{10}P_{{\rm X}}$ in experiment.

The minimization of the total deviation with respect to the model parameters is performed by the Broyden--Fletcher--Goldfarb--Shanno
(BFGS) quasi-Newton algorithm \citep{Fletcher-1987-Optimization}, as implemented in SciPy standard Python library.
This algorithm relies on the smoothness of a function to minimize, which is not the case in this work since the underlying dependence of
thermodynamic parameters on the cluster size, Eq. (\ref{eq:K-kink_H}), is piecewise-linear. In particular, we noticed that this non-smoothness
sometimes comprises the BFGS ability to find a minimum. To remedy that, we additionally convolute the piecewise-linear function with
the normalized gaussian $(2\pi\sigma^2)^{-1/2}e^{-n^2/2\sigma^2}$. Not too small $\sigma$ drastically increases the robustness of the BFGS convergence. On the other hand,  if $\sigma$ is small compared to one, then the resulting function is almost the same in magnitude at all $n$, as the one before the convolution. We choose $\sigma=0.3$ in this work.

\begin{figure*}[!htb]
\centering
\includegraphics[width=0.75\paperwidth]{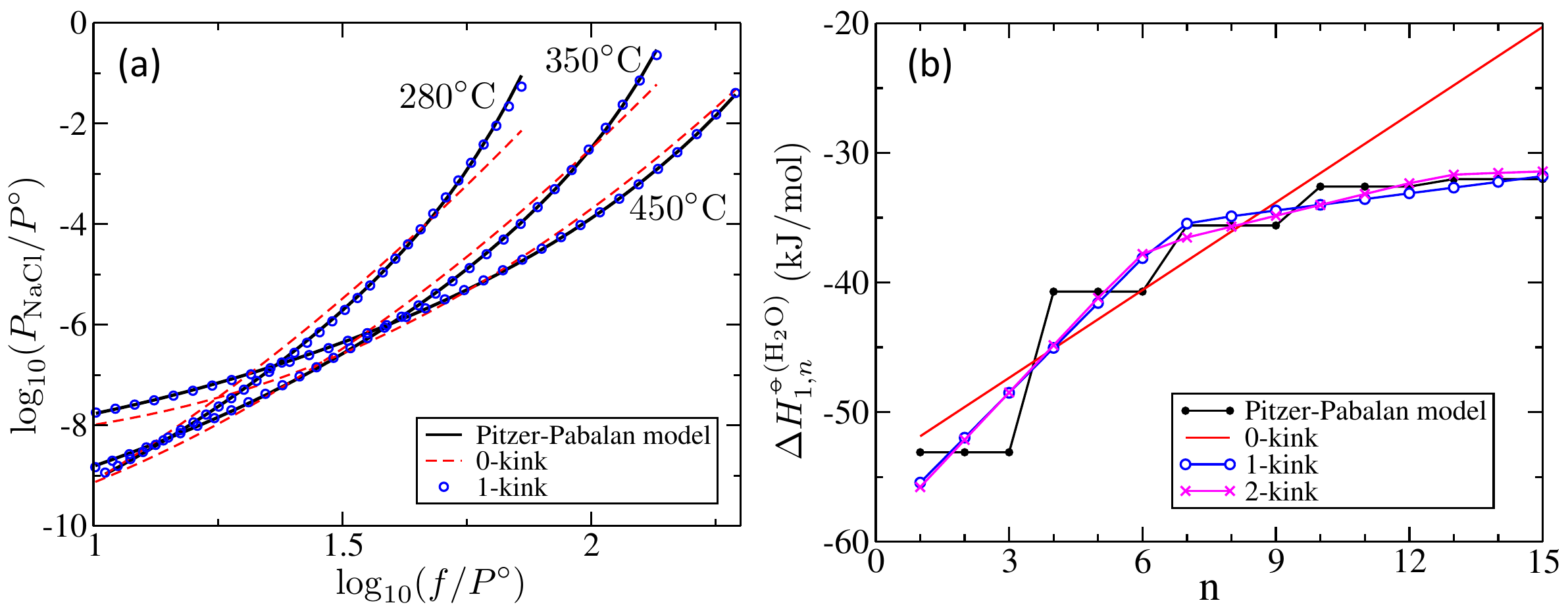}\caption{\label{fig:NaCl_fit_kinks}
Fitting of the partial pressure of the ideal gas of ${\rm NaCl}\text{:}({\rm H_2O})_n$ clusters, generated from the last three columns in Table \ref{tab:PP86_parms} at three different temperatures ($280^{\circ}{\rm C}$,
$350^{\circ}{\rm C}$ and $450^{\circ}{\rm C}$). Fitting
is performed using the continuous piecewise linear representation
of $\Delta H_{1,n}^{\stst({\rm H_{2}O})}$ as a function of $n$,
whereas entropy and heat capacity of hydration are taken $n$-independent.
The dependence of partial pressure of ${\rm NaCl}$-bearing water
clusters on the water fugacity for the Pitzer-Pabalan model (black
lines) and resulting fitting with the few-kink model (dashed red lines,
blue circles) are given in panel (a). Enthalpy of hydration for the
Pitzer-Pabalan model (black line) and the few-kink models (red, blue,
magenta lines) are plotted in panel (b).}
\end{figure*}

Figure \ref{fig:NaCl_fit_kinks} shows the results of fitting the data, generated from the last three columns of Table \ref{tab:PP86_parms}  (black lines), using the 0-, 1- and 2-kink function for $\Delta H_{1,n}^{\stst({\rm H_{2}O})}$,
and $n$-independent values for $\Delta S_{1,n}^{\stst({\rm H_{2}O})}$
and $\Delta C_{1,n}^{\stst({\rm H_{2}O})}$.
First, we perform the fitting by the 0-kink function to represent
$\Delta H_{1,n}^{\stst({\rm H_{2}O)}}$. The results of this fitting
are plotted by dashed red lines in panel (a). The resulting dependence
of $\Delta H_{1,n}^{\stst({\rm H_{2}O)}}$ on $n$ is shown in Figure
\ref{fig:NaCl_fit_kinks}(b) by the solid red line. More specifically,
the parameters produced by fitting are compiled in Table \ref{tab:0-kink_NaCl}.
\begin{table*}[!htb]
\centering
\begin{tabular}{|c|c|c|c|}
\hline 
$\Delta S_{1,n}^{\stst({\rm H_{2}O)}}/R$  & $\Delta C_{1,n}^{\stst({\rm H_{2}O)}}/R$  & $\Delta H_{1,1}^{\stst({\rm H_{2}O)}}$ (kJ/mol) & $h_{1}$ (kJ/mol)\tabularnewline
\hline 
\hline 
$-10.85$ & $2.744$ & $-51.86$ & $2.256$\tabularnewline
\hline 
\end{tabular}\caption{\label{tab:0-kink_NaCl}The results for fitting ${\rm NaCl}$ solubility with the 0-kink representation of $\Delta H_{1,n}^{\stst({\rm H_{2}O})}$. }
\end{table*}
 As is seen, the fit quality is reasonable although the deviations
between the solid black and dashed red lines are quite noticeable
in Figure \ref{fig:NaCl_fit_kinks}(a). The resulting best value of
the deviation function is $-1.901$.

The results of fitting for the 1-kink enthalpy function are shown
by blue circles in Figure \ref{fig:NaCl_fit_kinks}(a). As is seen, the agreement with the
data generated from the Pitzer-Pabalan model is very good, as is also clear from the best value of the deviation
function: $-2.905$. This significant improvement of the fitting quality
when switching from the 0-kink to 1-kink representation for $\Delta H_{1,n}^{\stst({\rm H_{2}O)}}$
suggests that the 0-kink model results in underfitting. The resulting
dependence of $\Delta H_{1,n}^{\stst({\rm H_{2}O)}}$ on $n$ is plotted
as blue circles in panel (b). The resulting parameters are given in
Table \ref{tab:1-kink_NaCl}.
\begin{table*}[!htb]
\centering
\begin{tabular}{|c|c|c|c|c|c|}
\hline 
$\Delta S_{1,n}^{\stst({\rm H_{2}O)}}/R$  & $\Delta C_{1,n}^{\stst({\rm H_{2}O)}}/R$ & $\Delta H_{1,1}^{\stst({\rm H_{2}O)}}$ (kJ/mol) & $h_{1}$ (kJ/mol) & $n_{1}$ & $h_{2}$ (kJ/mol)\tabularnewline
\hline 
\hline 
$-10.98$ & $3.005$ & $-55.43$ & $3.463$ & $5.777$ & $0.442$\tabularnewline
\hline 
\end{tabular}\caption{\label{tab:1-kink_NaCl}The results for fitting ${\rm NaCl}$ solubility with the 1-kink representation of $\Delta H_{1,n}^{\stst({\rm H_{2}O})}$.}
\end{table*}
We observed that if we do not allow $\Delta C_{1,n}^{\stst({\rm H_{2}O)}}/R$
to vary and set it to be an arbitrary constant value within interval from 2
to 4, it changes the value of the
deviation function by $\approx0.05$ at most. This means that, similarly
to what we observed when fitting the thermodynamics of sublimation,
Table \ref{tab:sublimation}, $\Delta C_{1,n}^{\stst({\rm H_{2}O)}}$
cannot be reliably extracted from the fitting.  Equilibrium constants $K_n=e^{-\Delta \tilde{G}^\circ_{1,n}/RT}$, where $\Delta \tilde{G}^\circ_{1,n}(T)$ is evaluated using the parameters in Table \ref{tab:1-kink_NaCl} and the first numerical column in Table \ref{tab:sublimation}, are tabulated in Table \ref{tab:Keq_NaCl} for a range of $n$ and $T$.

Finally, we perform the fitting using the 2-kink function for enthalpy.
The result of this fitting is not shown in Figure \ref{fig:NaCl_fit_kinks}(a)
since it superimposes with the 1-kink results.
According to the discussion just above, the heat capacity is
not varied and set to $\Delta C_{1,n}^{\stst({\rm H_{2}O)}}/R=3$.
The best value of the deviation function reaches $-3.301$ in this
case. The enthalpy change is shown in panel (b) by magenta crosses.
It is seen that it closely follows the 1-kink enthalpy dependence
at low $n$ but then deviates slightly. The resulting parameters are given in Table \ref{tab:2-kink_NaCl}.
\begin{table*}[!htb]
\centering
\begin{tabular}{|c|c|c|c|c|c|c|c|}
\hline 
$\Delta S_{1,n}^{\stst({\rm H_{2}O)}}/R$ & $\Delta C_{1,n}^{\stst({\rm H_{2}O)}}/R$ & $\Delta H_{1,1}^{\stst({\rm H_{2}O)}}$ (kJ/mol) & $h_{1}$ (kJ/mol) & $n_{1}$ & $h_{2}$ (kJ/mol) & $n_{2}$ & $h_{3}$ (kJ/mol)\tabularnewline
\hline 
\hline 
-11.0 & $3$ (forced) & $-55.78$ & $3.648$ & $5.064$ & $0.838$ & $11.802$ & $0.107$\tabularnewline
\hline 
\end{tabular}\caption{\label{tab:2-kink_NaCl}The results for fitting ${\rm NaCl}$ solubility with the $2$-kink representation of $\Delta H_{1,n}^{\stst({\rm H_{2}O})}$. }
\end{table*}
For the 1-kink representation for $\Delta H_{1,n}^{\stst({\rm H_{2}O)}}$, where only 4 independent parameters are used, the agreement between
model calculations and the benchmark is almost perfect. Introduction
of the second kink does not visibly improve the quality of fit, implying
that fitting with the 2-kink function to represent $\Delta H_{1,1}^{\stst({\rm H_{2}O)}}$
constitutes an overfitting. The original Pitzer-Pabalan $\Delta H_{1,n}^{\stst({\rm H_{2}O)}}$
dependence on $n$ is defined by 5 independent parameters, as is clear
from Table \ref{tab:PP86_parms}. We thus conclude that the original
work represents a case of overfitting.

\begin{figure}[!htb]
\centering
\includegraphics[width=0.38\paperwidth]{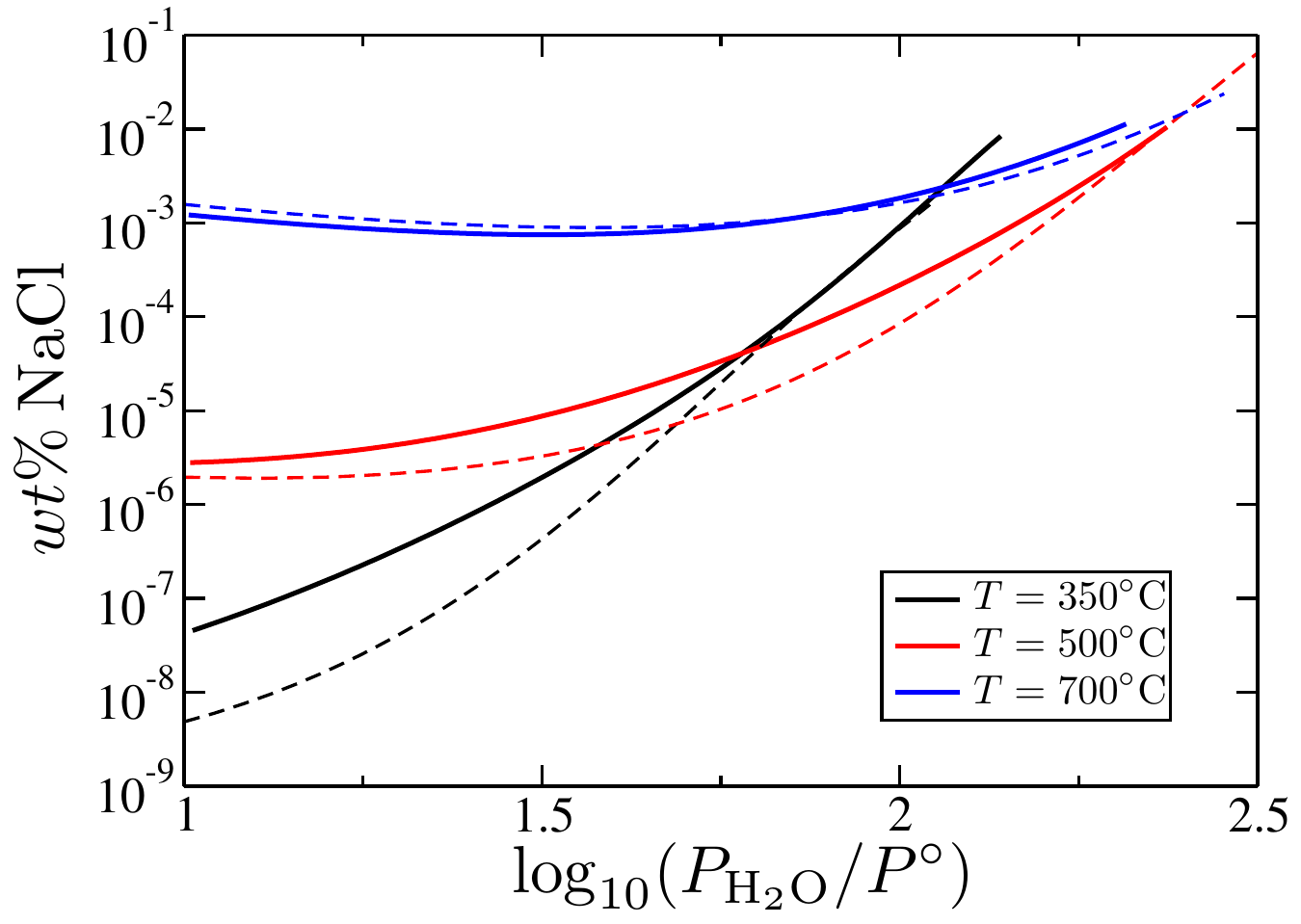}\caption{\label{fig:soWat_comp}
${\rm NaCl}$ wt\% (percentage by weight) versus
water vapor pressure for three different temperatures. Solid lines
are fitting results obtained in this work, Table \ref{tab:1-kink_NaCl}.
Dashed lines represent the results obtained from the the SoWat model
\citep{Driesner-2007-4880,Driesner-2007-4902}.}
\end{figure}

Finally, we want to compare our results to those obtained from a different
model describing the ${\rm NaCl}$ solubility in water vapor. Figure
\ref{fig:soWat_comp} shows the dependence of ${\rm NaCl}$ weight
percentage (wt\%) in water vapor as a function of vapor pressure at
three different temperatures.
Solid lines are obtained from Eq. (\ref{eq:P_tot}) using parameters
from Table \ref{tab:1-kink_NaCl}. Dashed lines are obtained for the vapor-solid equilibrium (liquid phase is absent) from the
SoWat computer code that implements the model to analyze the properties
of fluids and vapor in the ${\rm H_{2}O}$-${\rm NaCl}$ system \citep{Driesner-2007-4880,Driesner-2007-4902,SoWat-Driesner}.
As is seen, the agreement is satisfactory at higher temperatures,
where, importantly, our model involves extrapolation since it was calibrated using
the data only in the temperature range from $280^{\circ}{\rm C}$
to $450^{\circ}{\rm C}$, as is seen in Figure \ref{fig:NaCl_fit_kinks}(a). At lower temperatures, the SoWat model underestimates binding of water molecules to ${\rm NaCl}$ for smaller cluster sizes.

\section{Solubility of Copper Chloride in Water Vapor\label{sec:CuCl_solubility}}

\begin{figure*}[!htb]
\centering
\includegraphics[width=0.6\paperwidth]{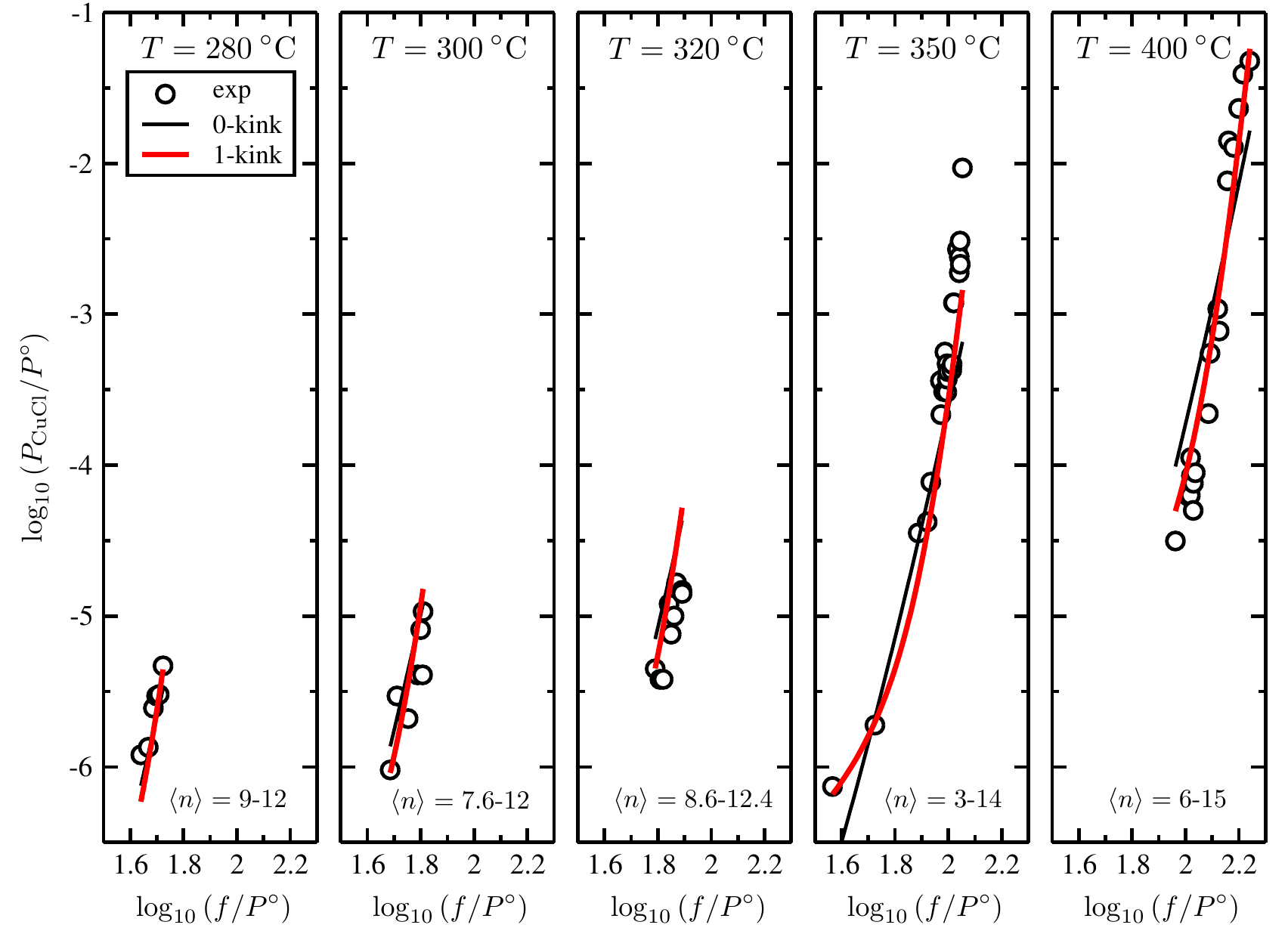}\caption{\label{fig:fit_CuCl}Partial pressure of ${\rm CuCl}$-bearing water
clusters versus the water fugacity for five different temperatures.
Experimental data from Ref. \citep{Migdisov-2014-33} is depicted
by black circles. Fitting results are plotted by the black and red
lines for the 0- and 1-kink models of $\Delta H_{1,n}^{\stst({\rm H_{2}O)}}$,
respectively.}
\end{figure*}

\begin{figure*}[!htb]
\centering
\includegraphics[width=0.8\paperwidth]{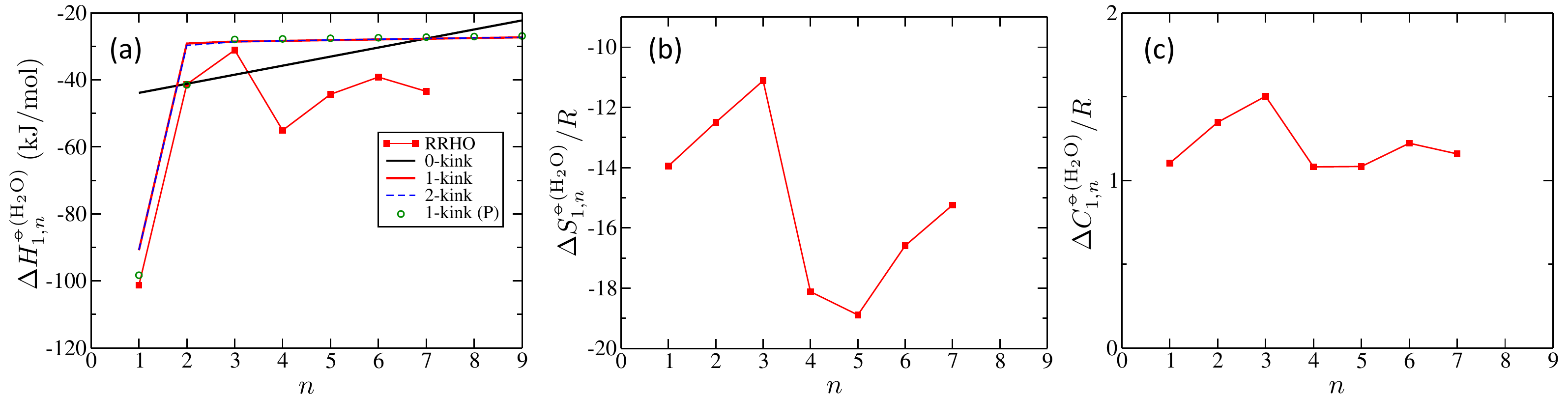}
\caption{\label{fig:dH_CuCl}(a) Standard molar enthalpy of hydration of ${\rm CuCl}$-bearing
clusters. Sold black, solid red and dashed blue lines correspond to
the first three numerical columns in Table \ref{tab:fit_CuCl}. Greens
circles correspond to the last column in Table \ref{tab:fit_CuCl}. Standard molar entropy and heat capacity of hydration are presented in panels (b) and (c), respectively.
Red squares in all the panels correspond to the results of the RRHO
calculations.}
\end{figure*}

The results of the fitting of the experimental $P_{{\rm CuCl}}$ dependence
on water fugacity $f$ for five different temperatures is plotted
in Figure \ref{fig:fit_CuCl}.
The black circles are the experimental data from Refs.
\citep{Archibald-2002-1611,Migdisov-2014-33}. To investigate the
degree of underfitting/overfitting we started, as before, by fitting
the experimental data with the 0-kink function for $\Delta H_{1,n}^{\stst({\rm H_{2}O)}}$,
taking the thermodynamics of sublimation from the third numerical
column of Table \ref{tab:sublimation}. Entropy of hydration $\Delta S_{1,n}^{\stst({\rm H_{2}O})}$ is assumed independent of $n$, and its value is varied in fitting. The heat capacity of hydration
is not varied and set to $\Delta C_{1,n}^{\stst({\rm H_{2}O})}/R=3$.
The resulting model $P_{{\rm CuCl}}(f)$ is plotted in Figure \ref{fig:fit_CuCl}
by the solid black lines. The resulting 0-kink function for $\Delta H_{1,n}^{\stst({\rm H_{2}O})}$
is plotted in Figure \ref{fig:dH_CuCl}(a) by the black line. 
The enthalpy change in the first hydration step, entropy and heat
capacity changes in any hydration step, as well as the resulting best
value of the total deviation, Eq. (\ref{eq:obj_func}), are given
in the first numerical column of Table \ref{tab:fit_CuCl}.
\begin{table*}[!htb]
\centering
\begin{tabular}{|c|c|c|c|c|c|}
\hline 
 & 0-kink & 1-kink & 2-kink & 1-kink & 1-kink (P)\tabularnewline
\hline 
\hline 
$\Delta H_{1,1}^{\stst({\rm CuCl})}$ (kJ/mol) & $-43.90$ & $-89.43$ & $-90.31$ & $-86.54$ & $-98.30$\tabularnewline
\hline 
$\Delta S_{1,n}^{\stst({\rm CuCl})}/R$ & $-9.198$ & $-9.719$ & $-9.732$ & $-9.331$ & $-9.684$\tabularnewline
\hline 
$\Delta C_{1,n}^{\stst({\rm CuCl})}/R$ & 3 & 3 & 3 & 1 & 3\tabularnewline
\hline 
$D$ & $-1.808$ & $-1.995$ & $-1.995$ & $-2.001$ & $-1.998$\tabularnewline
\hline 
\end{tabular}\caption{\label{tab:fit_CuCl}Results for the ${\rm CuCl}$ solubility fitting.
The fourth numerical column shows fitting results where the heat capacity
change was set to $\Delta C_{1,n}^{\stst({\rm H_{2}O})}/R=1$ instead
of usual $\Delta C_{1,n}^{\stst({\rm H_{2}O})}/R=3$ (the first three numerical columns). The last column shows the fitting results where the enthalpy
of hydration was represented by the same 1-kink form, but the sublimation
data was taken from Pankratz (the fourth numerical column in Table \ref{tab:sublimation}) and not from JANAF (the third numerical column) thermochemical tables. }
\end{table*}

We now re-run the fitting, with the only modification to the model
being the 1-kink, instead of 0-kink, function for the dependence of
$\Delta H_{1,n}^{\stst({\rm H_{2}O})}$ on $n$. The resulting thermodynamic
parameters, as well as the deviation, are plotted as the thick red
line in Figure \ref{fig:dH_CuCl}(a) and given in the second numerical
column of Table \ref{tab:fit_CuCl}. 
\begin{table*}[!htb]
\centering
\begin{tabular}{|c|c|c|c|c|c|}
\hline 
$\Delta S_{1,n}^{\stst({\rm H_{2}O)}}/R$  & $\Delta C_{1,n}^{\stst({\rm H_{2}O)}}/R$ & $\Delta H_{1,1}^{\stst({\rm H_{2}O)}}$ (kJ/mol) & $h_{1}$ (kJ/mol) & $n_{1}$ & $h_{2}$ (kJ/mol)\tabularnewline
\hline 
\hline 
 $-9.714$ & $3$ (forced)  & $-86.78$ & $60.86$  & $0.9575$  & 0.1915 \tabularnewline
\hline 
\end{tabular}\caption{\label{tab:1-kink_CuCl}The results of the fitting ${\rm CuCl}$ solubility with the 1-kink representation of $\Delta H_{1,n}^{\stst({\rm H_{2}O})}$.}
\end{table*}
The complete set of parameters extracted from the 1-kink fitting is given in Table \ref{tab:1-kink_CuCl}.
The corresponding equilibrium constants $K_n=e^{-\Delta \tilde{G}^\circ_{1,n}/RT}$, where $\Delta \tilde{G}^\circ_{1,n}(T)$ is evaluated using the parameters in Table \ref{tab:1-kink_CuCl} and the third numerical column in Table \ref{tab:sublimation}, are tabulated in Table \ref{tab:Keq_CuCl} for a range of $n$ and $T$.
The resulting model $P_{{\rm CuCl}}(f)$
is plotted by the red lines in Figure \ref{fig:fit_CuCl}. Comparing
these red lines with the 0-kink results (black lines) reveals much
better agreement with the experimental data at low water fugacities.
This visibly better agreement is also reflected quantitatively in
lower total deviation $D$ in Table \ref{tab:fit_CuCl}. Therefore,
similarly to what we had for ${\rm NaCl}$, the 0-kink model results
in underfitting. Switching to the 2-kink representation for $\Delta H_{1,n}^{\stst({\rm H_{2}O})}$
does not improve the total deviation for the assumed number of significant digits, as is seen in the third
numerical column of Table \ref{tab:fit_CuCl}. The resultant model
curves for $P_{{\rm CuCl}}(f)$ in the 2-kink case are not plotted
in Figure \ref{fig:fit_CuCl}, since they would practically coincide
with those generated in the 1-kink case. Comparing the thick red and
dashed blue lines in Figure \ref{fig:dH_CuCl} we see that all the 2-kink
fitting improvement is only a marginal smoothing of the $\Delta H_{1,n}^{\stst({\rm H_{2}O})}$
dependence on $n$ at $n\sim 2$. Therefore, we conclude that
the temperature range and scattering in the adopted experimental data
for ${\rm CuCl}$ allows one to extract more accurate dependence of
$\Delta H_{1,n}^{\stst({\rm H_{2}O})}$ on $n$ than that encoded by a 0-kink function. On the other hand, assuming the 2-kink model results in an overfit.
This is the reason why only the parameters for the 1-kink representation of the hydration enthalpy of ${\rm CuCl}$ are explicitly given in this work, Table \ref{tab:fit_CuCl}.

Up to this point, we have used the fixed value of the heat capacity
of hydration set to $\Delta C_{1,n}^{\stst({\rm H_{2}O})}/R=3$. To
explore the sensitivity of the fitting quality on the value of this
parameter, we also perform a 1-kink fitting with the heat capacity
set to $\Delta C_{1,n}^{\stst({\rm H_{2}O})}/R=1$. The resulting
hydration parameters and total deviation are given in the fourth numerical
column of Table \ref{tab:fit_CuCl}. As is seen, the total deviation
changes only marginally, and the model $P_{{\rm CuCl}}(f)$ curve,
if plotted in Figure \ref{fig:fit_CuCl}, would practically coincide
with those for the 1-kink case discussed before. Therefore, $\Delta C_{1,n}^{\stst({\rm H_{2}O})}$
can not be reliably extracted from the adopted experimental data.
We choose to set the heat capacity of hydration to $\Delta C_{1,n}^{\stst({\rm H_{2}O})}/R=3$
for definiteness.

There is a spread in ${\rm CuCl}$ sublimation parameters
in the literature, illustrated in this work by the difference between
the JANAF and Pankratz data, the third and fourth numerical columns
in Table \ref{tab:sublimation}. Re-fitting the experimental data assuming
Pankratz data for ${\rm CuCl}$ sublimation results in a 1-kink dependence
of $\Delta H_{1,n}^{\stst({\rm H_{2}O})}$ on $n$ shown by the green
circles in Figure \ref{fig:dH_CuCl}, and the fifth numerical column
in Table \ref{tab:fit_CuCl}. We see that the fit effectively compensates
the increase of the sublimation enthalpy, when going from JANAF to
Pankratz sublimation data, by the respective decrease of the enthalpy
of hydration when forming ${\rm CuCl}\text{:}{\rm H_{2}O}$. This
compensation is possible because what enters Eq. (\ref{eq:Px}) is
not the sublimation or hydration parameters separately, but only through
$\Delta\tilde{G}^{\circ}_{1,n}=\Delta G_{s}^{\circ}+\Delta G_{1,1}^{\circ({\rm H_{2}O})}+\Delta G_{1,2}^{\circ({\rm H_{2}O})}+...+\Delta G_{1,n}^{\circ({\rm H_{2}O})}$.
Accordingly, if the experimental data does not extend to very low
water fugacities where the contribution of anhydrous ${\rm CuCl}$
monomers to $P_{{\rm CuCl}}$ is dominant, fitting the experimental
data with Eq. (\ref{eq:Px}) can not provide, for example, $\Delta H_{s}^{\stst}$
and $\Delta H_{1,1}^{\stst({\rm H_{2}O})}$ separately. Evaluating
Eq. (\ref{eq:mean_n}) in the context of the red curves in Figure \ref{fig:fit_CuCl},
one can find that the lowest mean cluster size is around $\langle n\rangle\approx2.3$,
achieved at the lowest fugacity at $T=350\,^{\circ}{\rm C}$. In particular,
this means that clusters smaller that ${\rm CuCl}\text{:}\left({\rm H_{2}O}\right)_{2}$
do not contribute significantly to the the partial pressure of the
salt-bearing water clusters in the adopted experimental data, so no
fitting is able to separately provide $\Delta H_{s}^{\stst}$, $\Delta H_{1,1}^{\stst({\rm H_{2}O})}$
and $\Delta H_{1,2}^{\stst({\rm H_{2}O})}$. This is precisely why we
use the thermochemical tables that provide $\Delta G_{s}^{\circ}$
independently.

Results for our RRHO calculations for enthalpy, entropy and heat capacity
of hydration, calculated at $T_{{\rm ref}}$, are plotted by red squares
in Figure \ref{fig:dH_CuCl}. Similar to the ${\rm NaCl}$ case, RRHO
systematically underestimates the enthalpy and entropy of hydration,
compared to those obtained from fitting, if the clusters are not too
small. We expect that this is at least partially due to accounting
for only a single lowest-energy cluster of a given stoichiometry $(1,n)$.
In principle, Eq. (\ref{eq:react_h2o}) can be used with the understanding
that the both initial and final clusters are each in fact thermodynamic
ensembles of clusters of different configurations but the same stoichiometry.
Each such ensemble has larger enthalpy and larger entropy than the
lowest-energy cluster from this ensemble. The difference between the
thermodynamic ensemble and the lowest-energy cluster from this ensemble
should be the least pronounced for smaller clusters, where the significant
energy separation between the lowest and next-to-lowest configurations
may prevent the latter from contributing significantly to each $(1,n)$
ensemble. In other words, the smaller the cluster, the better the
corresponding thermodynamic ensemble is represented by the lowest-energy
cluster from this ensemble. We therefore expect that the net result
of accounting only for the lowest-energy cluster in Eq. (\ref{eq:react_h2o})
is the underestimation of enthalpy and entropy of hydration at high
fugacities.
In particular, since we take only the lowest-energy cluster
and not the ensemble of clusters of the same stoichiometry we underestimate
the enthalpy of the initial and final cluster in Eq. (\ref{eq:react_h2o}).
Not accounting for all the clusters of a given stoichiometry
underestimates the entropy.

\begin{figure}[!htb]
\centering
\includegraphics[width=0.38\paperwidth]{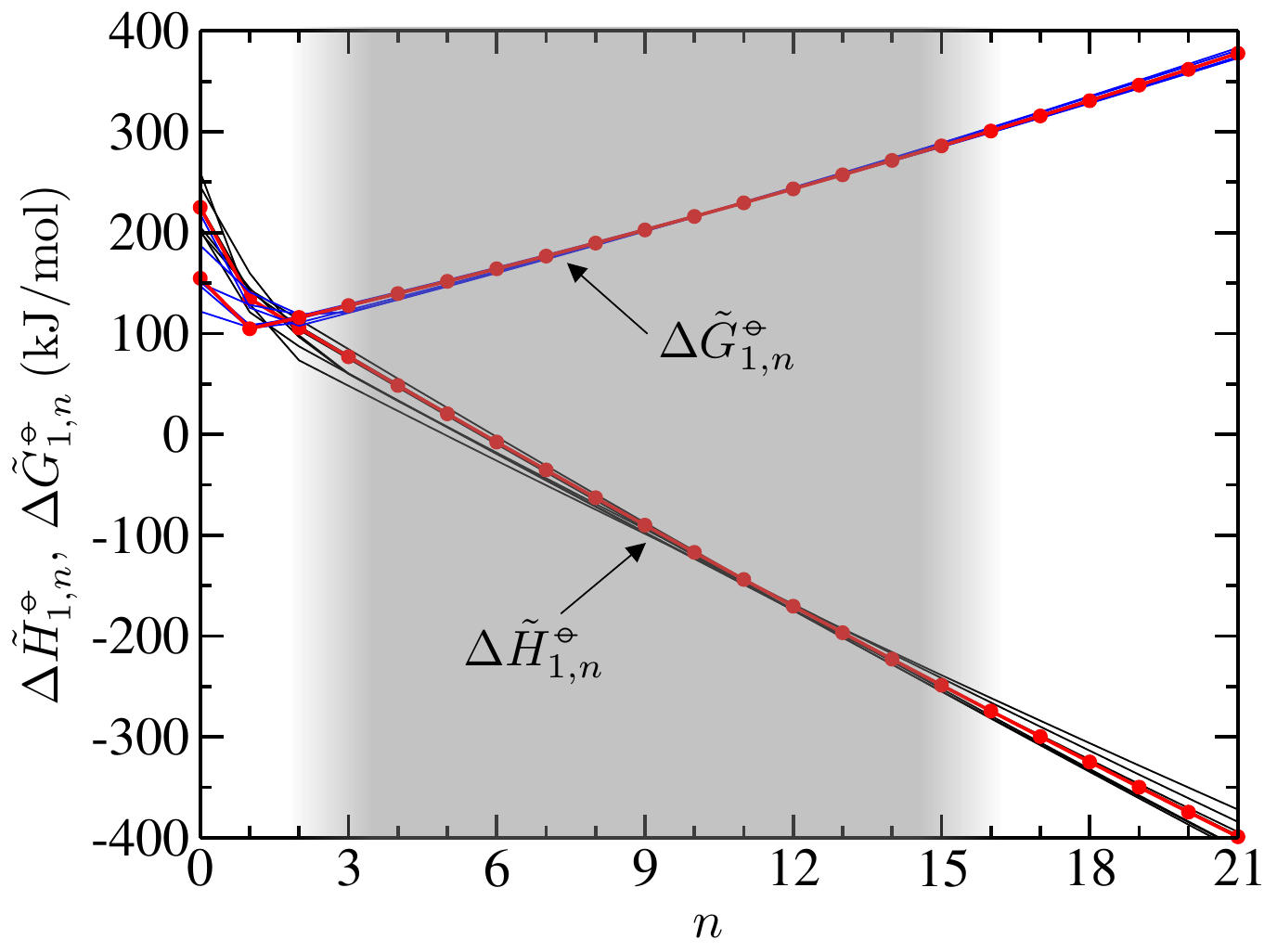}\caption{\label{fig:dGdH_cumulative_CuCl}The change of the standard molar
enthalpy and Gibbs free energy for the process ${\rm CuCl(cr)}+n{\rm H_{2}O}\rightarrow{\rm CuCl}\text{:}({\rm H_{2}O})_{n}$
at $T=T_{{\rm ref}}$. The black and blue lines are obtained by allowing
$\Delta H_{s}^{\stst}$ and $\Delta S_{s}^{\stst}$ to vary when fitting.
The red dots represent the results of the previous 1-kink fitting,
Figure \ref{fig:dH_CuCl}(a), where the sublimation parameters were
not varied and set to the values in the third numerical column of
Table \ref{tab:sublimation}.}
\end{figure}

To this point, fitting was performed while constraining the sublimation parameters to those extracted
from the thermochemical tables. This is not always possible for other
materials due to the lack of reliable thermochemical data. To demonstrate what happens if sublimation
parameters are not available, ${\rm CuCl}$ experimental data (circles in Figure \ref{fig:fit_CuCl}) was fit assuming $\Delta H_{s}^{\stst}$ and $\Delta S_{s}^{\stst}$ are unconstrained. Heat capacity of sublimation is still taken as constrained
since, as we saw above, heat capacities are related to weak second-order
deviations of Gibbs free energy with respect to temperature and, as
such, are not recovered reliably from fitting. Again, we set the heat
capacity of sublimation to $\Delta C_{s}^{\stst}/R=-3$ - a typical
value for sublimation, as can be seen from Table \ref{tab:sublimation}.
The results of such fitting, presented as the enthalpy $\Delta\tilde{H}_{1,n}^{\stst}$ and Gibbs free
energy $\Delta\tilde{G}_{1,n}^{\stst}$ of formation of a $(1,n$)-cluster formed out of $n$
gaseous water molecules and a single ${\rm CuCl}$ monomer extracted
from the crystalline bulk, are plotted in Figure \ref{fig:dGdH_cumulative_CuCl} as the thin black and blue lines, respectively.
 Multiple black and blue lines are obtained by varying initial values
of the fitting parameters and then performing BFGS optimization. Values
of the deviation function in the range from $-1.99$ to $-2.01$ are
reached for all the plotted black lines and the resulting $P_{{\rm CuCl}}(f)$
lays on top of the red lines in Figure \ref{fig:fit_CuCl}. It is seen
that the black (or blue) lines, corresponding to different initial
values of fitting parameters fed to BFGS before starting the minimization,
do not coincide. They also do not coincide with the red dots, which
represent the previous 1-kink fitting results where the sublimation
parameters were not varied and taken from the third numerical column
of Table \ref{tab:sublimation}. This means that the available experimental
data is not sufficient to uniquely constrain the newly introduced
fitting parameters, i.e., those of sublimation. In particular, since
$\Delta\tilde{H}_{1,0}^{\stst}=\Delta H_{s}^{\stst}$, we see that
the extracted enthalpy of sublimation varies in the range from $200$
to $260$ kJ/mol. Slopes of extracted $\Delta\tilde{H}_{1,n}^{\stst}$
are also varying suggesting that because of the uncertainty in $\Delta H_{s}^{\stst}$,
the enthalpies of hydration $\Delta\tilde{H}_{1,n}^{\stst({\rm H_{2}O})}$
can not be uniquely recovered.

Fitting the experimental data with Eq. (\ref{eq:Px}) allows one to evaluate the mean cluster size as a function of $f$ using Eq. (\ref{eq:mean_n}). Ranges of the mean cluster size, corresponding to the red lines in Figure \ref{fig:fit_CuCl}, are given in each panel
of that figure. As is seen, clusters of sizes smaller then 3 or large
than 15 are inaccessible in the experiment. The gray shading in Figure \ref{fig:dGdH_cumulative_CuCl}
highlights this range. Since the fitting is performed with Eq. (\ref{eq:Px}),
and of all the thermodynamic potentials it is $\Delta\tilde{G}_{1,n}^{\circ}$
that enters this expression directly, we expect the extracted values
of $\Delta\tilde{G}_{1,n}^{\circ}$ in the range $3\lesssim n\lesssim15$
to be most reliable. This is indeed the case in Figure \ref{fig:dGdH_cumulative_CuCl},
where blue lines practically superimpose within the gray shading area.
The reliability of the extracted temperature dependence of $\Delta\tilde{G}_{1,n}^{\circ}$,
expressed by values of $\Delta\tilde{H}_{1,n}^{\stst}$, $\Delta\tilde{S}_{1,n}^{\stst}$
and $\Delta\tilde{C}_{1,n}^{\stst}$, is expected to grow when the
experimentally accessible temperature range increases. The spread
in $\Delta\tilde{H}_{1,n}^{\stst}$ is seen to be larger than that
of $\Delta\tilde{G}_{1,n}^{\stst}$ in Figure \ref{fig:dGdH_cumulative_CuCl}
since the former does not enter Eq. (\ref{eq:Px}). However, $\Delta\tilde{H}_{1,n}^{\stst}$
still becomes relatively well defined, i.e., the spread is minimized,
in the middle of the $3\lesssim n\lesssim15$ range signifying that
$\frac{\partial\left(\Delta\tilde{G}_{1,n}^{\circ}\right)}{\partial T}=-\Delta\tilde{S}_{1,n}^{\circ}=\left(\Delta\tilde{G}_{1,n}^{\circ}-\Delta\tilde{H}_{1,n}^{\circ}\right)/T$
is extracted relatively reliably. The reliability of extraction of
the magnitude and temperature dependence of $\Delta\tilde{G}_{1,n}^{\circ}$
from fitting is thus directly related to, respectively, the range
of experimentally accessible water fugacities and temperatures.

\section{Discussion\label{sec:Discussion}}

In this section, we first discuss the chemical nature and strength of bonding in the ${\rm NaCl}\text{:}{\rm H_{2}O}$ and ${\rm CuCl}\text{:}{\rm H_{2}O}$ clusters. Then, in Sec. \ref{subsec:Large_n_enthalpy}, the large-cluster asymptotic behavior of the enthalpy of hydration is investigated. Finally, the assumption of temperature-independent heat capacity change $\Delta C^\circ$ is justified in Sec. \ref{subsec:dC_Tindep}.

\subsection{Nature of chemical bonding in ${\rm X}\text{\rm:}{\rm H_{2}O}$}

\begin{figure}[!htb]
\centering
\includegraphics[width=0.38\paperwidth]{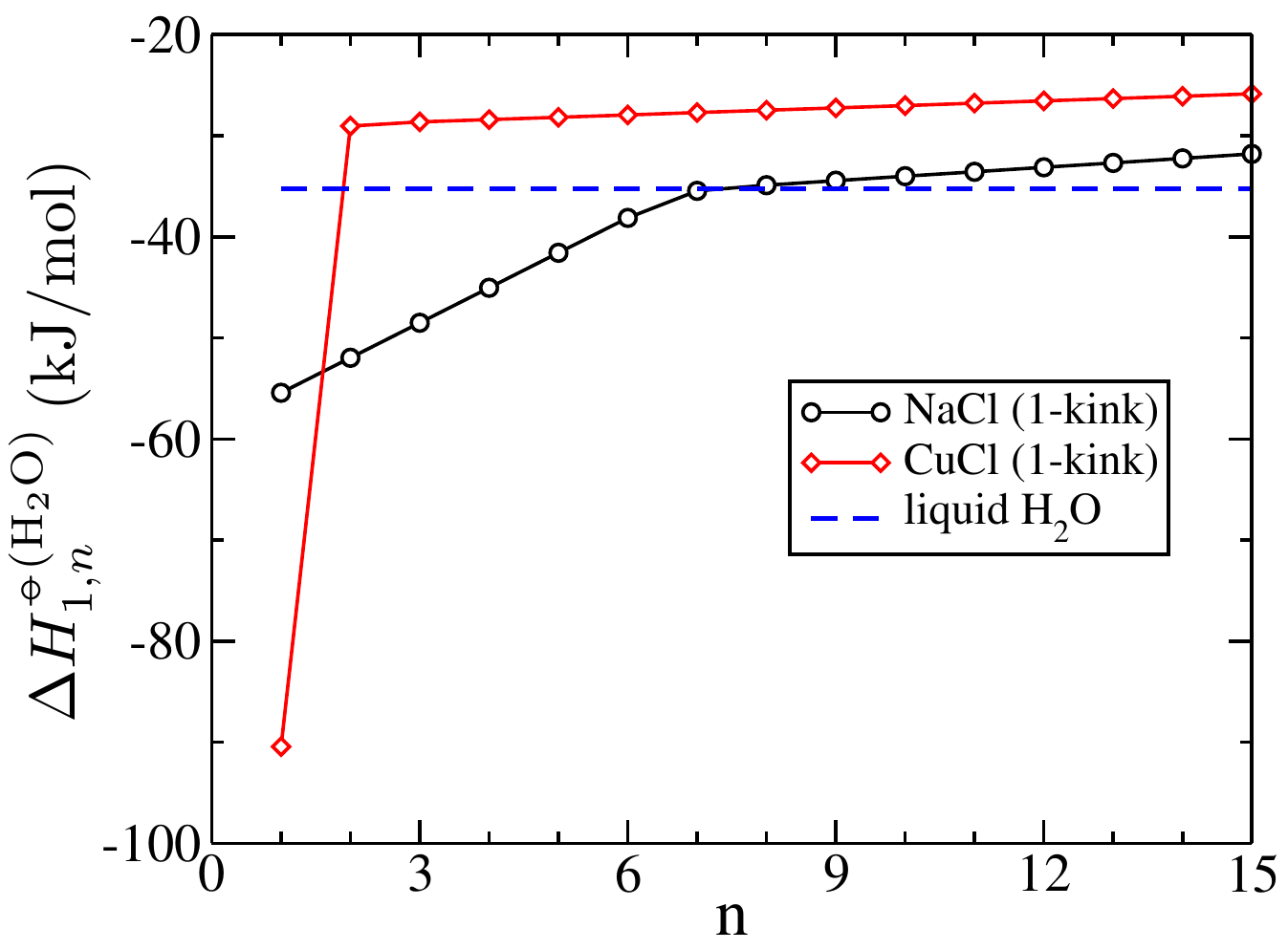}\caption{\label{fig:dH_NaCl_CuCl}Enthalpy of hydration of ${\rm NaCl}\text{:}({\rm H_{2}O})_{n}$
clusters (black circles) and ${\rm CuCl}\text{:}({\rm H_{2}O})_{n}$
clusters (red diamonds), replotted from Figures \ref{fig:NaCl_fit_kinks}(b)
and \ref{fig:dH_CuCl}(a), respectively. The dashed blue line is $\Delta H_{0,\infty}^{\stst({\rm H_{2}O})}$
from Table \ref{tab:water_clusters}.}
\end{figure}

Figure \ref{fig:dH_NaCl_CuCl} shows the dependence of the enthalpy of hydration on cluster size, where the enthalpies of hydration for ${\rm NaCl}$ (black circles) and ${\rm CuCl}$ (red diamonds) are reproduced from Figures
\ref{fig:NaCl_fit_kinks}(b) and \ref{fig:dH_CuCl}(a), respectively.
As is seen, the dependence of $\Delta H_{1,n}^{\stst({\rm H_{2}O})}$
on $n$ is qualitatively different for these two salts: (i) the enthalpy
of the first hydration $\Delta H_{1,1}^{\stst({\rm H_{2}O})}$ is
significantly higher in magnitude for ${\rm CuCl}$ than for ${\rm NaCl}$,
and (ii) the dependence of $\Delta H_{1,n}^{\stst({\rm H_{2}O})}$
on $n$ is smoother for ${\rm NaCl}$. We believe that these substantial
differences arise from the chemically different nature of interaction
of a water molecule with a salt monomer. The configurations of ${\rm NaCl}\text{:}{\rm H_{2}O}$
and ${\rm CuCl}\text{:}{\rm H_{2}O}$ clusters, as obtained from quantum
chemical calculations, are drawn in Figures \ref{fig:atomistics}(c)
and (d), respectively. The interaction of the water molecule with
the ${\rm NaCl}$ monomer is seen to result in a cluster where the
oxygen is attracted to ${\rm Na}$, and hydrogen to ${\rm Cl}$, resulting
in a typical dipole-dipole complex. On the other hand, oxygen binds
to ${\rm Cu}$ in the copper chloride monomer via the coordinate covalent
bond where the lone electron pair of oxygen is donated to the empty
electronic orbital of the copper. This covalent bonding results in
a linear configuration of the ${\rm CuCl}\text{:}{\rm H_{2}O}$ cluster
and the significantly larger magnitude of the enthalpy of first hydration,
compared to that for sodium chloride. The observed linear configuration
is consistent with previous observations \citep{Persson-2010-1901,Mei-2018-1}.

The first hydration of an anhydrous ${\rm CuCl}$ molecule saturates the electron
configuration of copper, so all the subsequent hydration steps, just
like all the hydration steps of ${\rm NaCl}$, proceed through the dipole-dipole
bonding of water molecules to a cluster. Our quantum chemical calculations with the Gaussian 16 software
package yield the dipole moments of $9.0\,{\rm D}$, $5.1\,{\rm D}$
and $2.2\,{\rm D}$ for the isolated sodium chloride, copper chloride
and water molecules, respectively. This means that the strength of
dipole-dipole bonding is expected to be larger for the ${\rm NaCl}$-based
complexes than for ${\rm CuCl}$-based ones. This is the reason why
the magnitude of $\Delta H_{1,n}^{\stst({\rm H_{2}O})}$ is noticeably higher
for the sodium chloride than for copper chloride at $n\geq2$. This difference is also reflected in the lowest-energy geometries of the ${\rm X}\text{:}({\rm H_2O})_2$ clusters in Figure \ref{fig:atomistics}, where the second water molecule binds directly to the salt molecule in panel (c), whereas the water-water hydrogen bond is formed in panel (d).  

\subsection{Strength of the dipole-dipole interaction in ${\rm NaCl}\text{:}{\rm H_{2}O}$}

Here, we explore whether the value of the enthalpy of
hydration $\Delta H_{1,1}^{\stst({\rm H_{2}O})}$ for ${\rm NaCl}$, obtained from the fitting,
is physically reasonable. In Sections \ref{sec:H2O_therm} and \ref{sec:X_therm}
we evaluated the thermodynamics of formation of ${\rm H_{2}O}$ and
${\rm NaCl}$ dimers, respectively. Since both of those dimers are
bound by the dipole-dipole interaction this information can be used
to estimate the enthalpy of formation of the ${\rm NaCl}\text{:}{\rm H_{2}O}$
cluster, which is also bound by the dipole-dipole interaction. The change of molar internal energy upon
the formation of a dipole-dipole bond between species ${\rm A}$ and
${\rm B}$ is approximately \citep{Jackson-1998}
\begin{equation}
\Delta E_{{\rm AB}}=E_{{\rm AB}}-E_{{\rm A}}-E_{{\rm B}}=-C\frac{d_{{\rm A}}d_{{\rm B}}}{r_{{\rm AB}}^{3}},
\end{equation}
where $d_{{\rm A}}$, $d_{{\rm B}}$ are the electric dipole moments
of the species, and $r_{{\rm AB}}$ is the characteristic distance
between them in the ${\rm AB}$ cluster. The constant prefactor $C$
will cancel out from the final expression for the enthalpy. The molar
enthalpy of the ideal gas is $H^{\stst}=E^{\stst}+RT_{{\rm ref}}$
and, therefore, we have
\begin{equation}
\Delta H_{{\rm AB}}^{\stst}=H_{AB}^{\stst}-H_{A}^{\stst}-H_{B}^{\stst}=-C\frac{d_{{\rm A}}d_{{\rm B}}}{r_{{\rm AB}}^{3}}-RT_{{\rm ref}}.
\end{equation}
This equation is inverted to obtain the characteristic distance between
the two molecules in a dimer
\begin{equation}
r_{{\rm AB}}=\left[-\frac{\Delta H_{{\rm AB}}^{\stst}+RT_{{\rm ref}}}{Cd_{{\rm A}}d_{{\rm B}}}\right]^{-1/3}.
\end{equation}
Using this approach, we express $r_{({\rm H_{2}O)_{2}}}$ and $r_{({\rm NaCl})_{2}}$
as functions of $\Delta H_{0,2}^{\stst({\rm H_{2}O)}}$ and $\Delta H_{2,0}^{\stst({\rm NaCl)}}$,
respectively, and then estimate the characteristic distance
between the molecules in the ${\rm NaCl}\text{:}{\rm H_{2}O}$ cluster
as a simple average of the inter-molecular distances in the two homogeneous
dimers, i.e., $r_{{\rm NaCl}\text{:}{\rm H_{2}O}}=(r_{({\rm NaCl})_{2}}+r_{({\rm H_{2}O})_2})/2$.
The enthalpy of formation of the ${\rm NaCl}\text{:}{\rm H_{2}O}$
cluster is then
\begin{equation}
\Delta H_{{\rm 1,1}}^{\stst({\rm H_{2}O})}=-C\frac{d_{{\rm NaCl}}d_{{\rm H_{2}O}}}{r_{{\rm NaCl}\text{:}{\rm H_{2}O}}^{3}}-RT_{{\rm ref}}.
\end{equation}
The constant prefactor $C$ cancels out and, by taking $\Delta H_{0,2}^{\stst({\rm H_{2}O)}}=-16.59\,{\rm kJ/mol}$
from Table \ref{tab:water_clusters} and $\Delta H_{2,0}^{\stst({\rm NaCl)}}\approx-200\,{\rm kJ/mol}$
from Figure \ref{fig:NaCldimer_form}(a), we obtain $\Delta H_{1,1}^{\stst({\rm H_{2}O})}=-53.5\,{\rm kJ/mol}$,
which is very close to the value of $-55.43\,{\rm kJ/mol}$ obtained
from the fitting and reported in Table \ref{tab:1-kink_NaCl}. We thus
see the value obtained in the fitting is reasonable on the grounds
of simple crude estimates based on the physics of the dipole-dipole
interaction.

\subsection{Large-$n$ behavior of the enthalpy of hydration \label{subsec:Large_n_enthalpy}}

The thermodynamics of hydration of a very large cluster ${\rm X}\text{:}({\rm H_{2}O})_{n}$
is not expected to be sensitive to the presence of the monomer ${\rm X}$ within it. More accurately, we approximate a large
$(1,n)$-cluster as a droplet of continuous dielectric liquid with
a point dipole at its center, representing ${\rm X}$. The optimal
shape of the droplet is generally non-spherical and determined by
the interplay between the surface tension and the magnitude of the
dipole, the latter favoring non-sphericity through the Keesom interaction
\citep{Atkins-2006-Physical}. However, the surface tension dominates
for the sufficiently large droplets resulting in a spherical shape
\citep{Shchekin-2002-318}. Molar internal energy of an ideal gas
of such droplets can be approximated as \citep{Machlin-2007-Aspects}
\begin{equation}
E_{1,n}^{\stst}\approx E_{{\rm liq}}n+\sigma n^{2/3},
\end{equation}
where $E_{{\rm liq}}$ is the internal energy corresponding to a mole
of water molecules residing in the bulk liquid water and $\sigma>0$
represents the positive surface energy of the droplet ($n^{2/3}$
is proportional to the droplet surface area). The corresponding enthalpy
is $H_{1,n}^{\stst}=E_{1,n}^{\stst}+RT_{{\rm ref}}$ and so the standard
enthalpy of hydration of the large clusters is
\begin{align}
\Delta H_{1,n}^{\stst({\rm H_{2}O})}&=H_{1,n}^{\stst}-H_{1,n-1}^{\stst}-H_{0,1}^{\stst}\nonumber\\
&\approx E_{{\rm liq}}-H_{0,1}^{\stst}+\frac{2}{3}\sigma/n^{1/3}.
\end{align}
One has $\Delta H_{1,\infty}^{\stst({\rm H_{2}O})}=E_{{\rm liq}}-H_{0,1}^{\stst}$
and so
\begin{equation}
\Delta H_{1,n}^{\stst({\rm H_{2}O})}=\Delta H_{1,\infty}^{\stst({\rm H_{2}O})}+\frac{2}{3}\sigma/n^{1/3}.
\end{equation}
An important observation here is that, since $\sigma>0$, $\Delta H_{1,n}^{\stst({\rm H_{2}O})}$
has to converge to $\Delta H_{1,\infty}^{\stst({\rm H_{2}O})}$ from above at large cluster sizes. However, when the clusters are small $\Delta H_{1,n}^{\stst({\rm H_{2}O})}<\Delta H_{1,\infty}^{\stst({\rm H_{2}O})}$.
It is thus expected that when the cluster size grows starting from
$n=1$, $\Delta H_{1,n}^{\stst({\rm H_{2}O})}$ first grows and overshoots
$\Delta H_{1,\infty}^{\stst({\rm H_{2}O})}$ and then decays
towards $\Delta H_{1,\infty}^{\stst({\rm H_{2}O})}$ resulting in
a non-monotonic behavior with respect to $n$. It is thus not surprising
that the fitting results in Figure \ref{fig:dH_NaCl_CuCl} overshoot
the expected $n\rightarrow\infty$ asymptotics plotted by the dashed blue line.

\subsection{Magnitude and temperature dependence of $\Delta C^\circ$ \label{subsec:dC_Tindep}}

The model, developed in this work, allows one to fit the experimental dependence of $P_{\rm X}$ on the water fugacity $f$ and temperature with Eq. (\ref{eq:Px}). The thermodynamics of formation of $(1,n)$-clusters enters this expression only through $\Delta \tilde{G}^\circ_{1,n}$. The temperature dependence of this Gibbs free energy change can be written as a Taylor series with respect to $T-T_{{\rm ref}}$. In particular, the magnitudes of the second- and third-order terms are determined by the change in heat capacity $\Delta \tilde{C}^\circ_{1,n}$ and its temperature derivative, respectively, evaluated at $T=T_{\rm ref}$. These two terms only weakly (especially the latter) affect the temperature variation of $\Delta \tilde{G}^\circ_{1,n}$, and, therefore, of $P_{\rm X}$, if the range of temperature variation is not too wide. Under these conditions, it is impossible to reliably extract the change in heat capacity and its temperature derivative from fitting. Exactly that was seen in Sec. \ref{subsec:Sublimation}, where fitting of the sublimation data for ${\rm CuCl}$ did not allow for reliable extraction of $\Delta \tilde{C}^\circ_{s}$. Similarly, fitting the ${\rm NaCl}$ solubility with the 1-kink model in Sec. \ref{subsec:fitting} only constrained the magnitude of $\Delta C^{\stst ({\rm H_2 O})}_{1,n}$ to within a factor of $\sim 2$. Since even the magnitude of $\Delta C^{\circ}$ is hard to extract reliably, the task of extracting its temperature derivative is unrealistic. We therefore assume that the temperature derivative is zero, which yields Eq. (\ref{eq:dG_1n_T}). This assumption is further supported by the observation that the heat capacities do not vary too much within the experimentally relevant temperature range in Figures \ref{fig:water_CvCp} and \ref{fig:NaCl_Cp}. Furthermore, since sublimation is the reverse of adding a salt monomer to a large pure-salt cluster, i.e., $\Delta G_{s}^{\circ}(T)=-\Delta G_{\infty,0}^{\circ}(T)$,
the temperature dependence of sublimation thermodynamics is effectively represented by the blue lines in Figure \ref{fig:NaCldimer_form}. Specifically, it is seen in Figure \ref{fig:NaCldimer_form}(c) that $\Delta C_{\infty,0}^{\circ}$ and therefore $\Delta C_{s}^{\circ}$, is a slowly varying function of temperature.

\section{Conclusion\label{sec:Conclusion}}

In this work we have developed a rigorous procedure for fitting
experimental data for solubility of salts in water vapor. The
procedure is based on the semi-empirical Pitzer-Pabalan model \citep{Pitzer-1986-1445}
and allows for extraction of thermodynamic parameters of formation
of salt-bearing water clusters. As an example, we applied the procedure to the solubility of ${\rm NaCl}$
and ${\rm CuCl}$ in water vapor, obtained elsewhere \citep{Pitzer-1986-1445,Migdisov-2014-33}.
Reliability of the fitting was determined by controlling the degree of underfitting/overfitting. The extracted
thermodynamic parameters are physically reasonable and, in particular,
provide insight into the nature of bonding between
anhydrous monomers of ${\rm NaCl}$ or ${\rm CuCl}$ and a single water
molecule. More specifically, the magnitude of the enthalpy for this first
hydration step was found to be significantly larger for ${\rm CuCl}$
than for ${\rm NaCl}$, which was rationalized by the formation of
the coordinate covalent bond in the former case and dipole-dipole
bond in the latter case. This has been confirmed by the quantum chemical
computations. We believe that this methodology can now be
applied to analyze the the data for solubility of solid salts in water vapor for many other materials.
In addition to solubility of solids in low-density water (i.e., vapor), the methods described here can also be applied to develop more comprehensive equations of state describing both phase equilibrium and volumetric (P-V-T) properties of water-salt systems \citep{Anderko-1993-1657}.

In this work, we focused on the so-called ``two-phase equilibrium'' where the vapor of water and salt molecules was in equilibrium with the crystalline salt in the absence of the liquid phase. The resulting thermodynamic parameters can now be used to evaluate the solubility of crystalline salt in water vapor at arbitrary temperature and pressure. A different type of experimental data, ``three-phase equilibrium'', where liquid water is present in addition to vapor and solid salt, is also available in literature \citep{Bischoff-1986-1437}. According to Gibbs' phase rule, the addition of liquid water to the vapor-solid system is simply introducing an extra constraint to the system so the pressure and temperature cannot be considered independent anymore. However, once the pressure and temperature are established for the vapor-liquid-solid equilibrium, the concentration of salt in vapor can still be calculated using Eq. (\ref{eq:Px}), so the results of this work are directly applicable.  On the other hand, the results of the present work are not directly applicable in situations where there is a vapor-liquid equilibrium in the absence of solid salt \citep{Bischoff-1986-1437}, as one would need to additionally supply a model to evaluate the salt activity in liquid water as a function of pressure, temperature and the salt concentration.  

\section*{Acknowledgments}
This work was supported by Los Alamos National Laboratory (LANL) Directed
Research and Development funds. This research used resources provided
by the Los Alamos National Laboratory Institutional Computing Program,
which is supported by the U.S. Department of Energy National Nuclear
Security Administration under Contract No. 89233218CNA000001. K.A.V.
is grateful to Sergei Ivanov (Center for Integrated Nanotechnologies)
for the discussion on nature of bonding in clusters.

\bibliographystyle{unsrt}
\bibliographystyle{elsarticle-num}
\biboptions{sort&compress}
\bibliography{bib}

\appendix

\section{Equilibrium Constants}

In this section we give the numerical values for the common logarithm of the equilibrium constants for the reaction
${\rm X}({\rm cr})+n{\rm H_{2}O}\rightarrow{\rm X}\text{:}({\rm H_{2}O})_{n}$. Tables \ref{tab:Keq_NaCl} and \ref{tab:Keq_CuCl} give data for the sodium and copper chlorides, respectively.
Equilibrium constants are defined as $K_n=e^{-\Delta \tilde{G}^\circ_{1,n}/RT}$ and the values in the tables are $\log_{10}K_n(T)$.

\begin{table*}[!htb]
\centering
\begin{tabular}{|c| c c c c c c c c|}
\hline
\multirow{2}{*}{$n$} & \multicolumn{8}{c|}{Temperature ($^\circ$C)}\\
& 100& 150& 200& 250& 300& 350& 400& 450\\
\hline
\hline
0 & $\:-24.005\:$ & $\:-20.218\:$ & $\:-17.244\:$ & $\:-14.848\:$ & $\:-12.879\:$ & $\:-11.233\:$ & $\:-9.837\:$ & $\:-8.639\:$\\
\hline
1 & $\:-20.953\:$ & $\:-18.125\:$ & $\:-15.891\:$ & $\:-14.081\:$ & $\:-12.584\:$ & $\:-11.325\:$ & $\:-10.252\:$ & $\:-9.325\:$\\
\hline
2 & $\:-18.385\:$ & $\:-16.460\:$ & $\:-14.921\:$ & $\:-13.660\:$ & $\:-12.605\:$ & $\:-11.709\:$ & $\:-10.935\:$ & $\:-10.261\:$\\
\hline
3 & $\:-16.302\:$ & $\:-15.222\:$ & $\:-14.333\:$ & $\:-13.584\:$ & $\:-12.942\:$ & $\:-12.382\:$ & $\:-11.888\:$ & $\:-11.447\:$\\
\hline
4 & $\:-14.704\:$ & $\:-14.411\:$ & $\:-14.127\:$ & $\:-13.854\:$ & $\:-13.594\:$ & $\:-13.346\:$ & $\:-13.109\:$ & $\:-12.883\:$\\
\hline
5 & $\:-13.591\:$ & $\:-14.028\:$ & $\:-14.304\:$ & $\:-14.470\:$ & $\:-14.562\:$ & $\:-14.600\:$ & $\:-14.599\:$ & $\:-14.570\:$\\
\hline
6 & $\:-12.962\:$ & $\:-14.072\:$ & $\:-14.862\:$ & $\:-15.432\:$ & $\:-15.845\:$ & $\:-16.144\:$ & $\:-16.358\:$ & $\:-16.507\:$\\
\hline
7 & $\:-12.708\:$ & $\:-14.447\:$ & $\:-15.716\:$ & $\:-16.660\:$ & $\:-17.372\:$ & $\:-17.912\:$ & $\:-18.323\:$ & $\:-18.636\:$\\
\hline
8 & $\:-12.532\:$ & $\:-14.890\:$ & $\:-16.632\:$ & $\:-17.945\:$ & $\:-18.950\:$ & $\:-19.727\:$ & $\:-20.333\:$ & $\:-20.806\:$\\
\hline
9 & $\:-12.417\:$ & $\:-15.388\:$ & $\:-17.596\:$ & $\:-19.273\:$ & $\:-20.568\:$ & $\:-21.579\:$ & $\:-22.376\:$ & $\:-23.008\:$\\
\hline
10 & $\:-12.365\:$ & $\:-15.940\:$ & $\:-18.609\:$ & $\:-20.646\:$ & $\:-22.226\:$ & $\:-23.469\:$ & $\:-24.454\:$ & $\:-25.242\:$\\
\hline
11 & $\:-12.374\:$ & $\:-16.547\:$ & $\:-19.671\:$ & $\:-22.062\:$ & $\:-23.925\:$ & $\:-25.395\:$ & $\:-26.567\:$ & $\:-27.508\:$\\
\hline
12 & $\:-12.445\:$ & $\:-17.209\:$ & $\:-20.782\:$ & $\:-23.523\:$ & $\:-25.664\:$ & $\:-27.358\:$ & $\:-28.713\:$ & $\:-29.805\:$\\
\hline
13 & $\:-12.579\:$ & $\:-17.925\:$ & $\:-21.941\:$ & $\:-25.028\:$ & $\:-27.443\:$ & $\:-29.358\:$ & $\:-30.894\:$ & $\:-32.135\:$\\
\hline
14 & $\:-12.774\:$ & $\:-18.695\:$ & $\:-23.149\:$ & $\:-26.577\:$ & $\:-29.262\:$ & $\:-31.396\:$ & $\:-33.109\:$ & $\:-34.497\:$\\
\hline
15 & $\:-13.030\:$ & $\:-19.521\:$ & $\:-24.406\:$ & $\:-28.170\:$ & $\:-31.122\:$ & $\:-33.470\:$ & $\:-35.359\:$ & $\:-36.890\:$\\
\hline
16 & $\:-13.349\:$ & $\:-20.400\:$ & $\:-25.712\:$ & $\:-29.808\:$ & $\:-33.022\:$ & $\:-35.582\:$ & $\:-37.642\:$ & $\:-39.316\:$\\
\hline
17 & $\:-13.730\:$ & $\:-21.335\:$ & $\:-27.067\:$ & $\:-31.489\:$ & $\:-34.963\:$ & $\:-37.730\:$ & $\:-39.960\:$ & $\:-41.773\:$\\
\hline
18 & $\:-14.172\:$ & $\:-22.324\:$ & $\:-28.471\:$ & $\:-33.215\:$ & $\:-36.943\:$ & $\:-39.916\:$ & $\:-42.313\:$ & $\:-44.262\:$\\
\hline
19 & $\:-14.677\:$ & $\:-23.367\:$ & $\:-29.923\:$ & $\:-34.984\:$ & $\:-38.964\:$ & $\:-42.139\:$ & $\:-44.699\:$ & $\:-46.783\:$\\
\hline
20 & $\:-15.243\:$ & $\:-24.465\:$ & $\:-31.424\:$ & $\:-36.798\:$ & $\:-41.025\:$ & $\:-44.398\:$ & $\:-47.120\:$ & $\:-49.337\:$\\
\hline
21 & $\:-15.871\:$ & $\:-25.618\:$ & $\:-32.974\:$ & $\:-38.656\:$ & $\:-43.127\:$ & $\:-46.695\:$ & $\:-49.575\:$ & $\:-51.922\:$\\
\hline
22 & $\:-16.561\:$ & $\:-26.825\:$ & $\:-34.572\:$ & $\:-40.558\:$ & $\:-45.269\:$ & $\:-49.029\:$ & $\:-52.065\:$ & $\:-54.539\:$\\
\hline
23 & $\:-17.313\:$ & $\:-28.086\:$ & $\:-36.220\:$ & $\:-42.505\:$ & $\:-47.451\:$ & $\:-51.400\:$ & $\:-54.589\:$ & $\:-57.188\:$\\
\hline
24 & $\:-18.127\:$ & $\:-29.403\:$ & $\:-37.916\:$ & $\:-44.495\:$ & $\:-49.673\:$ & $\:-53.808\:$ & $\:-57.147\:$ & $\:-59.868\:$\\
\hline
25 & $\:-19.002\:$ & $\:-30.774\:$ & $\:-39.661\:$ & $\:-46.530\:$ & $\:-51.936\:$ & $\:-56.252\:$ & $\:-59.739\:$ & $\:-62.581\:$\\
\hline
\end{tabular}\caption{\label{tab:Keq_NaCl}Common logarithm ($\log_{10}$) of $K_n=e^{-\Delta \tilde{G}^\circ_{1,n}/RT}$ for ${\rm NaCl}$. The dependance of $\Delta \tilde{G}^\circ_{1,n}$ on $n$
and temperature is evaluated using the parameters in Table \ref{tab:1-kink_NaCl} and the first numerical column in Table \ref{tab:sublimation}.}
\end{table*}

\begin{table*}[!htb]
\centering
\begin{tabular}{|c| c c c c c c c c|}
\hline
\multirow{2}{*}{$n$} & \multicolumn{8}{c|}{Temperature ($^\circ$C)}\\
& 100& 150& 200& 250& 300& 350& 400& 450\\
\hline
\hline
0 & $\:-24.217\:$ & $\:-20.453\:$ & $\:-17.501\:$ & $\:-15.127\:$ & $\:-13.177\:$ & $\:-11.549\:$ & $\:-10.171\:$ & $\:-8.990\:$\\
\hline
1 & $\:-15.660\:$ & $\:-13.441\:$ & $\:-11.691\:$ & $\:-10.276\:$ & $\:-9.107\:$ & $\:-8.126\:$ & $\:-7.291\:$ & $\:-6.571\:$\\
\hline
2 & $\:-15.736\:$ & $\:-14.042\:$ & $\:-12.690\:$ & $\:-11.583\:$ & $\:-10.659\:$ & $\:-9.873\:$ & $\:-9.196\:$ & $\:-8.606\:$\\
\hline
3 & $\:-15.895\:$ & $\:-14.717\:$ & $\:-13.755\:$ & $\:-12.950\:$ & $\:-12.264\:$ & $\:-11.669\:$ & $\:-11.148\:$ & $\:-10.684\:$\\
\hline
4 & $\:-16.084\:$ & $\:-15.417\:$ & $\:-14.842\:$ & $\:-14.337\:$ & $\:-13.888\:$ & $\:-13.483\:$ & $\:-13.115\:$ & $\:-12.778\:$\\
\hline
5 & $\:-16.302\:$ & $\:-16.143\:$ & $\:-15.953\:$ & $\:-15.746\:$ & $\:-15.531\:$ & $\:-15.315\:$ & $\:-15.099\:$ & $\:-14.886\:$\\
\hline
6 & $\:-16.549\:$ & $\:-16.895\:$ & $\:-17.086\:$ & $\:-17.175\:$ & $\:-17.194\:$ & $\:-17.164\:$ & $\:-17.099\:$ & $\:-17.010\:$\\
\hline
7 & $\:-16.825\:$ & $\:-17.673\:$ & $\:-18.243\:$ & $\:-18.625\:$ & $\:-18.875\:$ & $\:-19.030\:$ & $\:-19.115\:$ & $\:-19.149\:$\\
\hline
8 & $\:-17.131\:$ & $\:-18.476\:$ & $\:-19.423\:$ & $\:-20.096\:$ & $\:-20.575\:$ & $\:-20.914\:$ & $\:-21.148\:$ & $\:-21.302\:$\\
\hline
9 & $\:-17.465\:$ & $\:-19.305\:$ & $\:-20.625\:$ & $\:-21.587\:$ & $\:-22.295\:$ & $\:-22.815\:$ & $\:-23.196\:$ & $\:-23.471\:$\\
\hline
10 & $\:-17.829\:$ & $\:-20.160\:$ & $\:-21.851\:$ & $\:-23.100\:$ & $\:-24.033\:$ & $\:-24.734\:$ & $\:-25.261\:$ & $\:-25.655\:$\\
\hline
11 & $\:-18.222\:$ & $\:-21.040\:$ & $\:-23.100\:$ & $\:-24.633\:$ & $\:-25.791\:$ & $\:-26.671\:$ & $\:-27.342\:$ & $\:-27.854\:$\\
\hline
12 & $\:-18.645\:$ & $\:-21.947\:$ & $\:-24.371\:$ & $\:-26.188\:$ & $\:-27.567\:$ & $\:-28.624\:$ & $\:-29.439\:$ & $\:-30.068\:$\\
\hline
13 & $\:-19.096\:$ & $\:-22.879\:$ & $\:-25.666\:$ & $\:-27.763\:$ & $\:-29.362\:$ & $\:-30.596\:$ & $\:-31.553\:$ & $\:-32.297\:$\\
\hline
14 & $\:-19.577\:$ & $\:-23.837\:$ & $\:-26.984\:$ & $\:-29.358\:$ & $\:-31.177\:$ & $\:-32.585\:$ & $\:-33.682\:$ & $\:-34.541\:$\\
\hline
15 & $\:-20.087\:$ & $\:-24.820\:$ & $\:-28.325\:$ & $\:-30.975\:$ & $\:-33.010\:$ & $\:-34.591\:$ & $\:-35.828\:$ & $\:-36.800\:$\\
\hline
16 & $\:-20.626\:$ & $\:-25.830\:$ & $\:-29.689\:$ & $\:-32.613\:$ & $\:-34.863\:$ & $\:-36.615\:$ & $\:-37.990\:$ & $\:-39.074\:$\\
\hline
17 & $\:-21.194\:$ & $\:-26.865\:$ & $\:-31.076\:$ & $\:-34.271\:$ & $\:-36.734\:$ & $\:-38.656\:$ & $\:-40.168\:$ & $\:-41.364\:$\\
\hline
18 & $\:-21.792\:$ & $\:-27.926\:$ & $\:-32.486\:$ & $\:-35.950\:$ & $\:-38.625\:$ & $\:-40.715\:$ & $\:-42.363\:$ & $\:-43.668\:$\\
\hline
19 & $\:-22.419\:$ & $\:-29.013\:$ & $\:-33.919\:$ & $\:-37.650\:$ & $\:-40.535\:$ & $\:-42.791\:$ & $\:-44.573\:$ & $\:-45.988\:$\\
\hline
20 & $\:-23.075\:$ & $\:-30.125\:$ & $\:-35.375\:$ & $\:-39.371\:$ & $\:-42.463\:$ & $\:-44.885\:$ & $\:-46.800\:$ & $\:-48.322\:$\\
\hline
21 & $\:-23.760\:$ & $\:-31.263\:$ & $\:-36.854\:$ & $\:-41.113\:$ & $\:-44.411\:$ & $\:-46.997\:$ & $\:-49.043\:$ & $\:-50.672\:$\\
\hline
22 & $\:-24.475\:$ & $\:-32.427\:$ & $\:-38.357\:$ & $\:-42.875\:$ & $\:-46.377\:$ & $\:-49.125\:$ & $\:-51.302\:$ & $\:-53.037\:$\\
\hline
23 & $\:-25.218\:$ & $\:-33.617\:$ & $\:-39.882\:$ & $\:-44.659\:$ & $\:-48.363\:$ & $\:-51.272\:$ & $\:-53.577\:$ & $\:-55.417\:$\\
\hline
24 & $\:-25.991\:$ & $\:-34.833\:$ & $\:-41.430\:$ & $\:-46.463\:$ & $\:-50.368\:$ & $\:-53.435\:$ & $\:-55.869\:$ & $\:-57.811\:$\\
\hline
25 & $\:-26.793\:$ & $\:-36.074\:$ & $\:-43.001\:$ & $\:-48.288\:$ & $\:-52.391\:$ & $\:-55.617\:$ & $\:-58.176\:$ & $\:-60.221\:$\\
\hline
\end{tabular}\caption{\label{tab:Keq_CuCl}Common logarithm ($\log_{10}$) of $K_n=e^{-\Delta \tilde{G}^\circ_{1,n}/RT}$ for ${\rm CuCl}$. The dependance of $\Delta \tilde{G}^\circ_{1,n}$ on $n$
and temperature is evaluated using the parameters in Table \ref{tab:1-kink_CuCl} and the third numerical column in Table \ref{tab:sublimation}.}
\end{table*}

\end{document}